\documentclass[%
 reprint,
 amsmath,amssymb,
 aps,
superscriptaddress
]{revtex4-2}

\usepackage{graphicx}
\usepackage{dcolumn}
\usepackage{bm}
\usepackage{hyperref}
\usepackage{xcolor}
\usepackage{multirow}
\usepackage{float}

\newcommand{\Da}{D_{\text{A}}}

\newcommand{\adata}{\alpha_{\text{data}}}
\newcommand{\dTb}{\delta T_{\text{b}}}
\newcommand{\xHI}{x_{\text{HI}}}
\newcommand{\amin}{\alpha_{\text{min}}}
\newcommand{\dsamp}{d_{\text{samp}}}
\newcommand{\rbubbles}{r_{\text{bubbles}}}
\newcommand{\Tcmb}{T_{\text{CMB}}}
\newcommand{\Ob}{\Omega_{\text{b}}}
\newcommand{\Om}{\Omega_{\text{m}}}
\newcommand{\hws}{h_{\text{ws}}}
\newcommand{\cMpc}{\text{cMpc}}

\newcommand{\cGpcSquare}{\text{cGpc}^2}
\newcommand{\cmfast}{\texttt{21cmFAST}}
\newcommand{\toolscm}{\texttt{tools21cm}}

\begin{document}


\title{An Alcock-Paczy\'nski Test on Reionization Bubbles for Cosmology}

\author{Emilie Thélie} 
\email{emilie.thelie@austin.utexas.edu}
\affiliation{Department of Astronomy, University of Texas at Austin \\
2512 Speedway, Austin, TX 78712, USA}

\author{Franco Del Balso}
\affiliation{Department of Physics and Trottier Space Institute, McGill University,  Montreal, QC, H3A 2T8, Canada}
\affiliation{Ciela, 
Computation and Astrophysical Data Analysis Institute \\ 
Montreal, QC, H3A 2T8, Canada}
 
 \author{Julian B. Mu\~noz} 
\affiliation{Department of Astronomy, University of Texas at Austin \\
2512 Speedway, Austin, TX 78712, USA}

\author{Adrian Liu} 
\affiliation{Department of Physics and Trottier Space Institute,  McGill University,  Montreal, QC, H3A 2T8, Canada}
\affiliation{Ciela, 
Computation and Astrophysical Data Analysis Institute \\ 
Montreal, QC, H3A 2T8, Canada}

\date{\today}

\begin{abstract}
In this paper, we propose an Alcock-Paczy\'nski (AP) test to constrain cosmology using HII bubbles during the Epoch of Reionization. Similarly to cosmic voids, a stack of HII bubbles is spherically symmetric because ionizing fronts propagate isotropically on average (even if individual bubbles may not be spherical), making them standard spheres to be used in an AP test.
Upcoming 21-cm observations, from the Square Kilometer Array (SKA) for instance, will contain tomographic information about HII regions during reionization. However, extracting the bubbles from this signal is made difficult because of instrumental noise and foreground systematics. Here, we use a neural network to reconstruct neutral-fraction boxes from the noisy 21-cm signal, from which we extract bubbles using a watershed algorithm.
We then run the purely geometrical AP test on these stacks, showing that a SKA-like experiment will be able to constrain the product of the angular-diameter distance $\Da$ and Hubble parameter $H$ at reionization redshifts with $\sim 2\%$ precision, robustly to astrophysical and cosmological uncertainties within the models tested here.
This AP test, whether performed on 21-cm observations or other large surveys of ionized bubbles, will allow us to fill the knowledge gap about the expansion rate of our Universe at reionization redshifts.

\end{abstract}

\maketitle


\section{Introduction} \label{sec:intro}

Current observations from galaxy surveys and the Cosmic Microwave Background (CMB) allow us to constrain cosmology at either low ($z\lesssim$ few) or very high redshifts ($z\sim1100$), leaving a gap in knowledge at intermediate times. 
Measurements of reionization-era ($z\sim 5-15$) observables such as the 21-cm line of hydrogen are opening a new promising window to this epoch \citep{Furlanetto2006,Pritchard2012,Liu2013,Bera2023}, though they often suffer from deep degeneracies between astrophysics and cosmology \citep{Greig2015,Park2019,Hassan2020,Hothi2024}.

Standard rulers and candles provide a well-tested way to extract cosmology robustly amidst astrophysics. 
Famous standard candles include cepheids or supernovae \citep{HST:2000azd,Rigault2015,Riess2016,Riess2019}, whereas standard rulers include the Baryon Acoustic Oscillation (BAO) feature of the spatial galaxy distribution \citep{BOSS2017,BOSS2017b,BOSS2017c}. 
These rulers are ``calibrated'' standard rulers, as we know their physical size (e.g., $\sim 150$ cMpc for the BAO sound horizon). 
A similar and promising probe of cosmology consists of using ``uncalibrated'' standard rulers, for which we do not know the size but we do know their shape. Cosmic voids are an example of  ``standard spheres'', as they have to be spherical on average \citep{Hamaus2020}. 
Both calibrated and uncalibrated standard rulers are robust to astrophysics and allow us to constrain cosmology, in particular the expansion history through the Hubble parameter $H$ and the comoving angular diameter distance $\Da$, by comparing their angular size on the sky and along the line of sight.

We currently lack standard rulers at intermediate redshifts. During the Cosmic Dawn and the Epoch of Reionization (EoR), at redshifts $z\sim5-30$, the 21-cm line will provide us with tomographic maps of those redshifts and thus potentially new standard rulers. This signal, emitted by the spin-flip transition of neutral hydrogen, is the target of radio telescopes such as the Hydrogen Epoch of Reionization Array (HERA; \citet{HERA2023}), the LOw Frequency ARray (LOFAR; \citet{Mertens2020}), the Murchison Widefield Array (MWA; \citet{Trott2020}), the Precision Array for Probing the Epoch of Re-ionization (PAPER; \citet{Parsons2010}, or the Long Wavelength Array (LWA; \citet{Dilullo2021}). 
\citet{Munoz2019} proposed that velocity-induced acoustic oscillations (VAOs) provide a new (calibrated) standard ruler that can constrain cosmology during Cosmic Dawn ($15\lesssim z \lesssim 20$; see also \cite{Sarkar2023,Munoz:2019rhi}).
During reionization, however, VAOs are expected to be weaker~\cite{Cain:2020npm,Park:2020ydt}, requiring new standard rulers.
Recently, \citet{Fronenberg2024} have shown that it is possible to probe the BAO scale at $2\lesssim z\lesssim14$ using the CMB and line intensity mapping.

In this paper, we propose a new kind of uncalibrated standard ruler to constrain cosmology at EoR redshifts ($5 \lesssim z \lesssim 15$) through an Alcock-Paczy\'nski (AP) test \citep{Alcock1979}. This test has been used in the context of cosmic voids (see e.g. \citet{Hamaus2020}), and relies purely on geometry. The key insight is that the cosmological principle ensures that a stack of a statistically large number of voids is spherically symmetric (even if each void is not). Therefore, when converting angular and redshift coordinates (with which we observe objects on the sky) to comoving distances, the stack should remain spherical. Since this conversion depends on the cosmology assumed at a given redshift (through $\Da$ and $H$)  only when assuming the true value of $\Da H$ for the conversion will the standard spheres remain spherical, hence testing cosmology. 
As the AP test is purely geometrical, it allows us to put constraints on $\Da H$ independently of astrophysics.

Here, we extend this concept to ionized bubbles during reionization. Similarly to cosmic voids, a stack of bubbles can also serve as an uncalibrated standard sphere.
During the EoR, galaxies emit UV radiation that is expected to ionize the intergalactic medium (IGM) around them, forming ``bubbles". These are large-scale (a few to tens of megaparsecs, see e.g. \cite{Furlanetto2004,Giri2019}) pockets of HII gas surrounding the first galaxies of our Universe, growing as reionization proceeds, merging with nearby bubbles, and ending by filling our entire Universe when reionization is complete.
Past work has studied their size distributions, morphology, and the topology of their complex network (see e.g. \citep{Miralda-Escude:1998adl,Nusser2005,McQuinn2007,Friedrich2011,Gazagnes2021,Thelie2022,Thelie2023,Jamieson2024}). In this paper, we show that the ionized bubbles are also a powerful tool for constraining cosmological parameters.

While theoretically promising, an AP test using ionized bubbles is an experimentally challenging prospect, as directly observing HII regions during the EoR is not trivial.
A very promising probe for such detections are the 21-cm tomographic maps that experiments like the Square Kilometer Array (SKA; \cite{Koopmans2015}) will observe in the next decade. 
As the 21-cm signal is emitted by neutral hydrogen, they can in principle directly map the ionized bubbles during reionization (bubbles corresponding to zero-signal regions). However, astrophysical foreground contamination is three to four orders of magnitude stronger than the cosmological signal, making its detection challenging \citep{Bernardi2009,Bernardi2010}. With interferometers such as the SKA, these foregrounds create a problem of mode mixing, and become localized in a wedged-shape region of the Fourier space called the ``foreground wedge", typically deemed unusable for cosmology \citep{Datta2010,Morales2010,Parsons2012,Vedantham2012,Trott2012,Hazelton2013,Pober2013,Thyagarajan2013,Liu2014,Liu2014b}.
Techniques involving deep-learning have recently been developed to reconstruct the Fourier modes residing in this foreground wedge \citep{Li2019,Makinen2021,GH2021,Bianco2024,Kennedy2024,Sabti2024,Bianco2024b,Beardsley:2014bea}. 
In this work, we use \citet{Kennedy2024}'s neural network to recover hydrogen neutral fraction maps from wedge-removed 21-cm maps, and use them to extract ionized bubbles using the watershed algorithm described by \citet{Lin2016}. We have chosen to run an AP test with bubbles obtained from wedge-removed 21-cm maps, but any way of detecting a statistical stack of bubbles from observations (see e.g. \cite{Lu2024,Mukherjee2024}) could also be used in future work.

The rest of this paper is organized as follows. We start by describing how an AP test works and how we implement one in this work, using a toy model as example, in Sec. \ref{sec:aptest}. In Sec. \ref{sec:unet_bubbles}, we explain how we obtain bubble stacks from 21-cm observations. Afterwards, our results are presented in Sec. \ref{sec:results}, before concluding in Sec. \ref{sec:conclusion}. 
In a series of appendices, we lay out some of the technical details of our approach as well as a series of robustness tests. Appendix \ref{app:z_lightcone_effects} studies different redshifts and the impact of using tomographic images of the 21-cm signal incorporating light-cone effects. Appendix \ref{app:varying_params} shows that our results are robust to reasonable parameter variations. Lastly, Appendix \ref{app:watershed_appendix} details the adaptations to the traditional watershed algorithm that are necessary for our proposed AP test.
In all this work, the \citet{PlanckCollab2018} cosmology is used as fiducial, and all distances are comoving unless otherwise indicated.

\section{Alcock-Paczy\'nski tests} \label{sec:aptest}

Standard rulers allow us to constrain the geometry and expansion history of the universe.
For instance, \citet{Eisenstein2005} first used the well-known physical scale that acoustic physics imprint on galaxy distributions, the BAO at 150 cMpc, to constrain cosmological parameters.
Even with a ruler of unknown length there is a way to learn cosmology. Instead of using astrophysical objects with a known distance scale (standard rulers), we use objects that have a known shape (standard spheres) to perform an AP test~\cite{Alcock1979}.
\citet{Hamaus2020} performed such a test on a stack of voids identified at low $z$ to measure $\Da H$ (though not $\Da$ or $H$ independently) at redshifts $z<0.6$.
In this paper we will extend this concept to higher $z$ by using ionized bubbles, extracted from 21-cm maps, to constrain the same product $\Da H$ at reionization redshifts.

In this Section, we explain how an AP test uses statistical isotropy to extract cosmology, and describe how we use the sphericity of ionized bubbles to constrain $\Da H$ before showing a toy model of this test.

\subsection{Principle} \label{sec:aptest_principle}

\begin{figure*}
\centering
\includegraphics[width=0.8\textwidth]{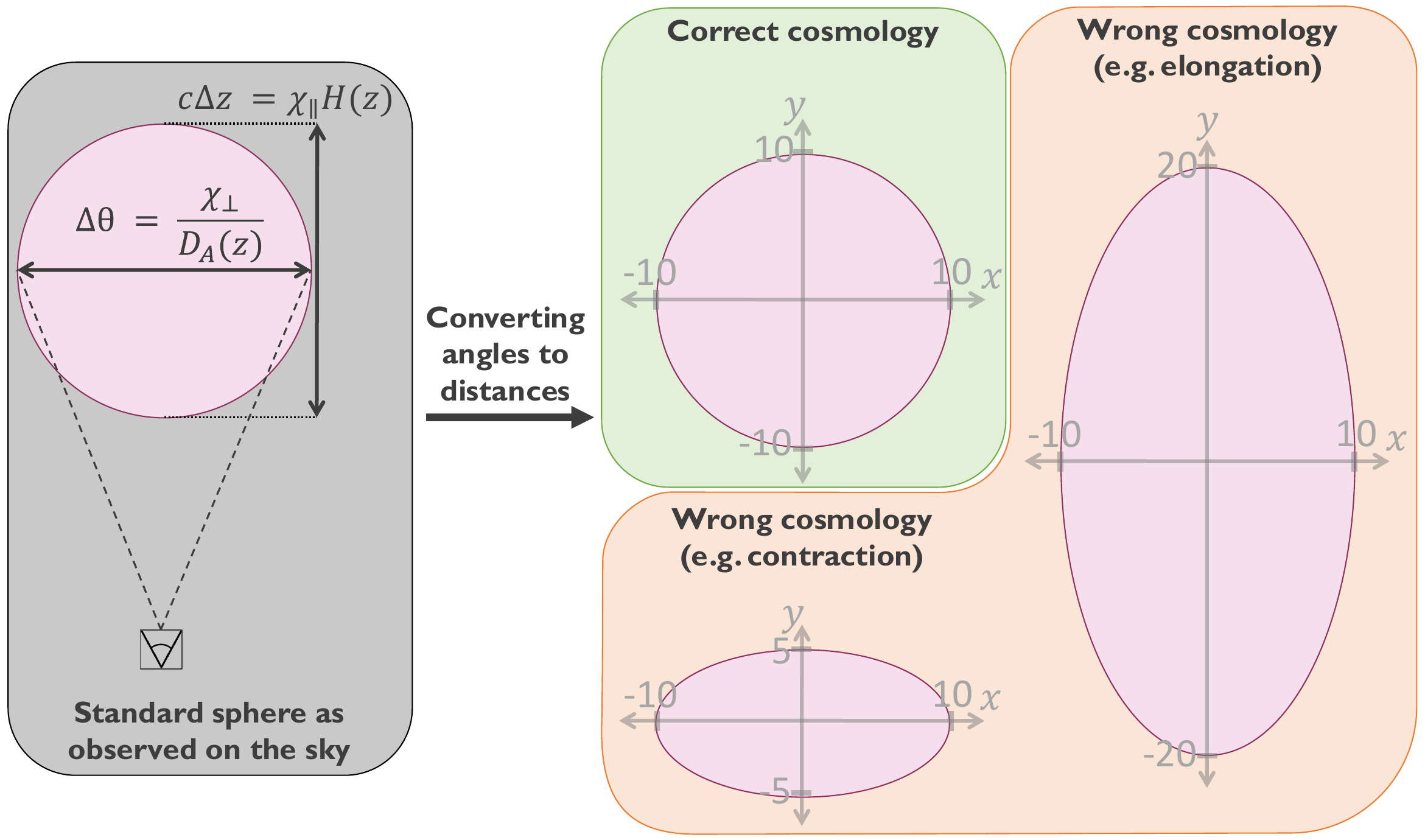}
\caption{Schematic of an AP test. In the gray box (left), we see a spherical object observed on the sky with angle and redshift coordinates. Converting these coordinates to comoving distances forces us to assume a cosmological model through $\Da$ and $H$. If we assumed the correct one, then our ``standard sphere'' will appear spherical in comoving coordinates, as in the green box. However, if the cosmological model is not the correct one, the object will be distorted along the line of sight, as in the orange box. This allows us to constrain the product $\Da\times H$.
\label{fig:ap_test_scheme}}
\end{figure*}

An AP test extracts the product $\Da H$ using the conversion from observed angles and redshifts to comoving distances, as illustrated in Fig.~\ref{fig:ap_test_scheme}. 
A ``standard sphere'' (e.g., a stack of voids or reionization bubbles, shown in the gray box of Fig.~\ref{fig:ap_test_scheme}) is spherical in comoving coordinates, which are related to the observed redshift $\delta z$ and  angle $\delta\theta$ separations through
\begin{equation}
    \begin{split}
        \chi_{\parallel} &= \dfrac{c \delta z}{H(z)} \\
        \chi_{\perp} &= \Da(z) \delta\theta \\
    \end{split}
\label{eq:comoving_distances}
\end{equation}
with $\chi_{\parallel}$ the comoving ``depth'' (i.e.~line of sight distance) of the object, $\chi_{\perp}$ its comoving ``width'' (or transverse distance), and $c$ the speed of light. 
Clearly the conversion from observable coordinates to comoving space, in which the objects are spherical, depends on the Hubble rate $H(z)$ and on the comoving angular diameter distance $\Da(z)$.
Since we do not know the underlying cosmology, we have to assume a fiducial $H'$ and $\Da'$ to obtain comoving distances:
\begin{equation}
    \begin{split}
        \chi_{\parallel}' &= \frac{H(z)}{H'(z)} \chi_{\parallel} = q_{\parallel}(z) \chi_{\parallel} \\
        \chi_{\perp}' &=  \frac{\Da'(z)}{\Da(z)} \chi_{\perp} = q_{\perp}(z) \chi_{\perp} \\
    \end{split}
\label{eq:cdist_to_fid}
\end{equation}
which are related to the true comoving distances by $q_{\parallel}\equiv H/H'$ and $q_{\perp}\equiv \Da'/\Da$. 

In the case of cosmic voids or reionization bubbles, we do not know their physical size, but isotropy ensures their shape: a stack of such objects should be spherical.
We therefore introduce the deformation parameter $\alpha$, which is also called the AP parameter \citep{Hamaus2020}:
\begin{equation}
    \alpha \equiv \frac{q_{\parallel}}{q_{\perp}} = \frac{\Da H}{(\Da H)'} ,
\label{eq:alpha}
\end{equation}
where all these quantities implicitly depend on redshift $z$. 
The true bubble stack is intrinsically spherical (i.e., $\chi_{\parallel} = \chi_{\perp}$), but the {\it observed} stack in the physical distances space will only appear spherical for a certain value of $\alpha$, corresponding to the product $\Da \times H$ of the correct underlying cosmology (as in the green box of Fig. \ref{fig:ap_test_scheme}). In contrast, the stack will be either contracted or stretched along the line of sight if the wrong cosmology (i.e. wrong $\Da' H' \neq \Da H$) is assumed (as in the orange box of Fig. \ref{fig:ap_test_scheme}).
Said differently, under the assumption of intrinsic sphericity, $\alpha$ can also be expressed as $\chi_\parallel^\prime / \chi_\perp^\prime$. It is thus a measurable parameter that can be computed using observable quantities only.

\subsection{A test of sphericity on bubbles} \label{sec:aptest_model}

Let us now describe how we apply an AP test to reionization bubbles. 
We assume that we have a stack $N$ of many bubbles obtained either from observations or simulations. The bubbles are centered on the same cell of a 3D Cartesian grid (which is the geometrical center of the bubbles), and  $N(x,y,z)$ measures how many are superimposed on each voxel of the grid.

In practice, we measure (RA, DEC, redshift), and need to assume a cosmology to obtain comoving coordinates ($x,y,z$). 
We will not know the ``true'' cosmology $\Da H$, so we will parameterize the mismatch between the ``true'' cosmology and a fiducial assumed cosmology with $\adata \equiv \Da H / (\Da H)'$. Our bubble stack therefore depends implicitly on $\adata$. 
In order to infer this parameter $\adata$, we need to look for a deformation of the bubble stack along the line of sight, which we set here to be the $z$ direction, relative to the transverse dimension. In other words, we want to test the sphericity of $N$.

Unlike standard rulers, which have a well-known distance scale (e.g., a radial feature in $N$), our uncalibrated standard ruler should be spherical but can have any radial profile. We will, then, compare each profile $N$ to its closest spherically symmetric stack, which we dub $\overline{N}(r=\sqrt{x^2+y^2+z^2})$, found by simply radially averaging $N(x,y,z)$.
We will vary the ``input'' deformation $\alpha$ as a free parameter by stretching (or contracting) the measured stack $N$ by an amount $\alpha$ along the line of sight ($z$), and for each input we find the closest spherically symmetric stack $\overline{N}_{\alpha}$ through
\begin{equation}
    \overline{N}_{\alpha}(x,y,z) = \overline{N}(r = \sqrt{x^2 + y^2 + (z/\alpha)^2}). 
\label{eq:RP_map}
\end{equation}
In our AP test, we therefore want to vary this $\alpha$ deformation to search when $N$ is most spherical (i.e., closest to $\overline{N}_{\alpha}$), which will be true for $\alpha=\alpha_{\rm data}$.

We can now describe how we use sphericity for this AP test. We work with the inclination angle $\theta=\arccos(z/r)$. In the case where a stack is spherical, all angles $\theta$ would be equally represented in a voxel histogram of $\theta$ weighted by the stack, making this histogram flat. But, if the stack is stretched or contracted along the line of sight, then the histogram would not be flat against $\theta$. 
We use this logic to see how far $N$ deviates from sphericity, and therefore compute a ratio of histograms of the inclination angle $\theta$ weighted by the maps $N$ and $\overline{N}_{\alpha}$ (divided by the Jacobian $r^2\sin{\theta}$ due to a change of coordinates from Euclidean to spherical). 
Taking the ratio of the $N$ histogram with the $\overline{N}_{\alpha}$ one allows this test to be independent on the radial profile of the bubble stack, while also removing some numerical effects due to sampling on a grid. 
Supposing the bubble stack is deformed by some $\adata$ value, the histogram ratio with respect to $\theta$ should be flat when $\alpha=\adata$.

However, as we are working with data sampled on a Cartesian grid, there are some $\theta$ values at which this sphericity test still creates artifacts (unwanted peaks near $\theta=0$, $\pi/2$ and $\pi$ in the histogram ratio). We have therefore chosen to not compute the histograms at those $\theta$ values, and work with the restrained $\theta$ interval $\mathcal{I}_{\theta}=[0.25,1.45]\cup[1.75,2.9]$.
Our sphericity test is hence the following:
\begin{equation}
    \forall \theta\in\mathcal{I}_{\theta}, \frac{H_{N}}{H_{\overline{N}_{\alpha}}}(\theta) = 1 \iff \adata=\alpha.
\label{eq:finding_adata}
\end{equation}

We can then infer the deformation $\adata$ in our data (and thus the value of $\Da H$) by minimizing the following $\chi^2$:
\begin{equation}
    \begin{split}
        \chi^2(\alpha) &= (\mathbf{D}-\mathbf{M})^T \mathbf{C}^{-1} (\mathbf{D}-\mathbf{M}) \\
        &= \sum_{i,j} \left(D_i-1 \right) C^{-1}_{ij} \left(D_j-1 \right),
    \end{split}
\label{eq:chi2cov}
\end{equation}
where $\mathbf{D}$, $\mathbf{M}$ and $\mathbf{C}$ are the data, model, and covariance matrices respectively. The data $D_i=(H_{N}/H_{\overline{N}_{\alpha}})(\theta_i)$ is the ratio of the histograms of $N$ and $\overline{N}_{\alpha}$ evaluated at an angle $\theta_i$, and the model $M_i$ is simply 1.
Given that this is a fairly complex observable of the input maps, we numerically compute $\mathbf{C}$ from simulated bubble stacks in order to capture the correlation between the $\theta$ bins of the histograms. In the case of the mock 21-cm observations we generate in this study, we use 400 realizations of bubble stacks to compute this matrix (see Sec. \ref{sec:unet_BS} for more details).

\subsection{Toy model} \label{sec:toymodel}

\begin{figure}
\centering
\includegraphics[width=0.5\textwidth]{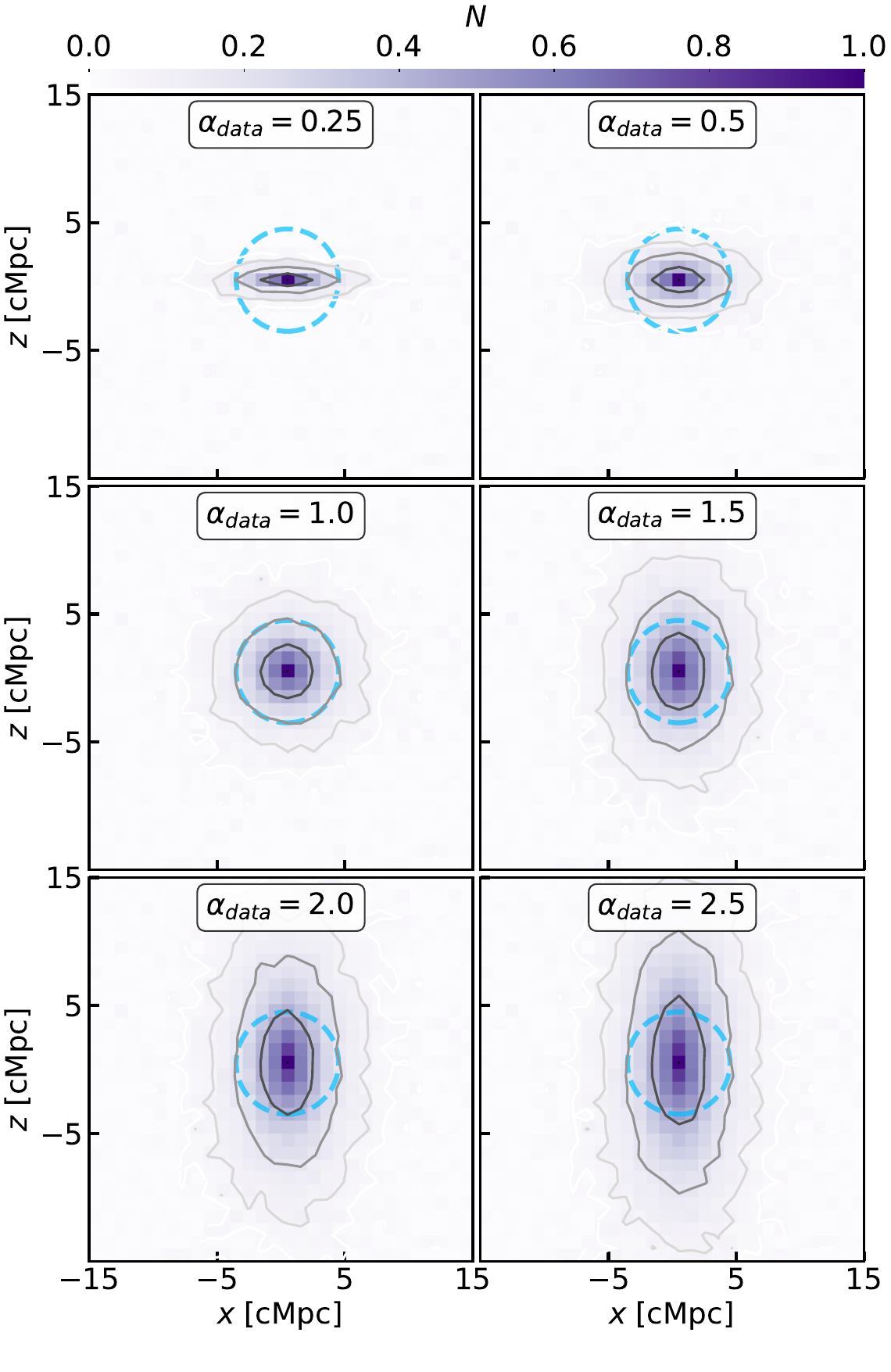}
\caption{Simulated bubble stacks for our toy case detailed in Sec. \ref{sec:toymodel}. In each panel, we show a universe with a different $\Da H$, and thus a different deformation $\adata$. Bubble stack isocontours are shown in greyscale, and a blue circle with radius $\approx 5$ cMpc has been added in every panel to guide the eye. In the $\adata=1$ case the assumed fiducial cosmology is the true cosmology, so the stacks appear spherical. In the other cases, the assumed fiducial cosmological is offset, producing a contraction for $\adata<1$ and an elongation for $\adata>1$.
\label{fig:toymodel_maps}}
\end{figure}

\begin{figure}
\centering
\includegraphics[width=0.5\textwidth]{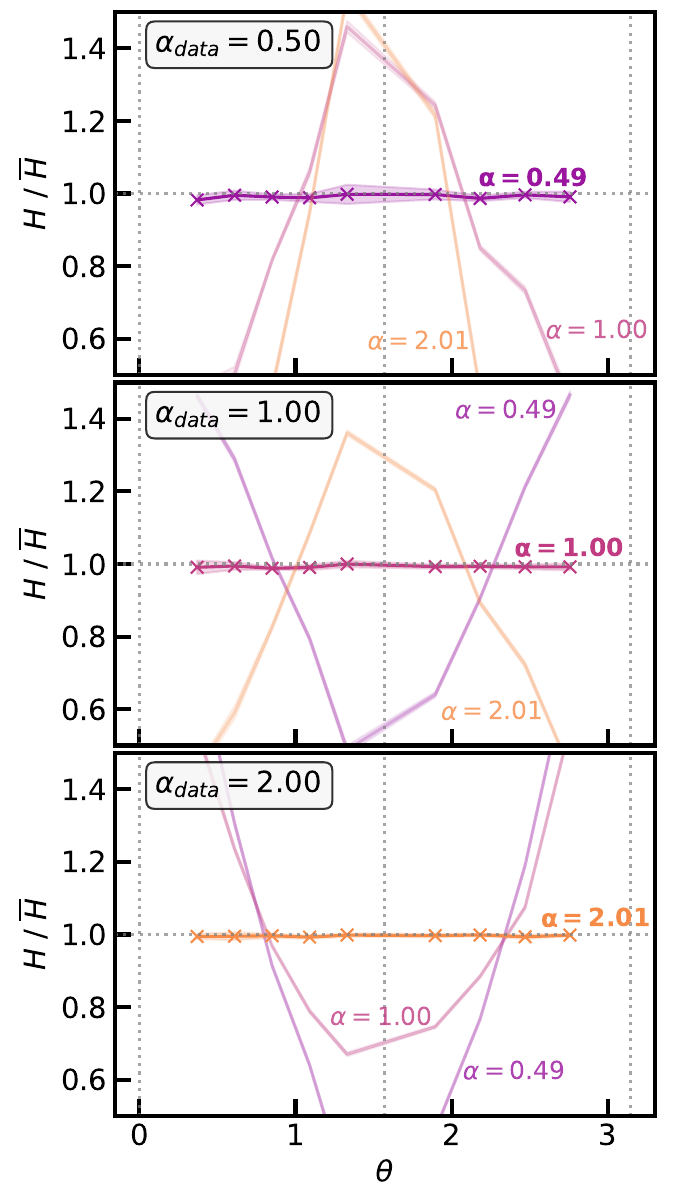}
\caption{Ratio of histograms $H/\bar{H}$ ($=H_{N}/H_{\overline{N}_{\alpha}}$) against line-of-sight angle $\theta$, used to test the sphericity of the toy models. Each panel shows mock data from a different cosmology in which the deformation $\adata$ is either 0.5, 1, or 2. The colors stand for the different values of the deformation parameter $\alpha$ of the model, and the shaded area represent the standard deviation calculated from 10 runs of each data set.
\label{fig:toymodel_PDFangles}}
\end{figure}

\begin{figure}
\centering
\includegraphics[width=0.5\textwidth]{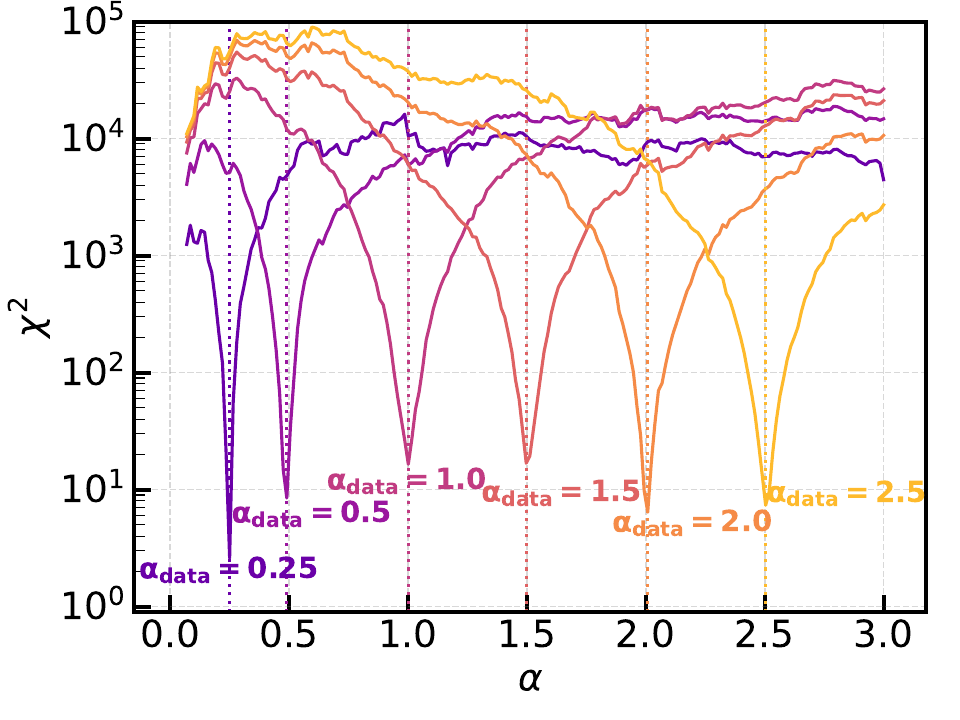}
\caption{$\chi^2$ of our AP test for the toy model. Each color represents a different universe in which the deformation $\adata$ is among the following values: 0.25, 0.5, 1, 1.5, 2, 2.5. The $\adata$ deformation can be inferred by looking at the value of $\alpha$ at which the $\chi^2$ is minimized, with errors given by the width of $\chi^2$ around that value.
In all cases tested here the deformation is correctly inferred within the error bars.
\label{fig:toymodel_chi2}}
\end{figure}

To illustrate our AP test and build intuition, we will start with a toy example, in which we make mock bubble stacks akin to the ones coming from reionization simulations, but with a well-known input profile. 
We generate them with the following radial profile:
\begin{equation}
    u(r) = e^{-\gamma r},
\label{eq:analytic_RP}
\end{equation}
where $\gamma=0.445$. This value of $\gamma$ was obtained doing a fit of a radial profile from a bubble stack generated with a $\cmfast$ simulation that was ran with fiducial parameters (version 3.0.3; \citet{Mesinger2011,Muray2020}; please refer to Sec. \ref{sec:unet_BS} for a description of how we construct a stack). Normalizing the radial profile to a certain number of bubbles is unnecessary here as only the bubble stack shape matters for the AP test.
The mock bubble stacks are constructed as
\begin{equation}
    N(x,y,z) = u(r_{\adata}) + N_{\text{noise}},
\label{eq:toymodel_RP_map}
\end{equation}
with $r_{\adata}^2=x^2+y^2+(z/\adata)^2$, and $N_{\text{noise}}$ a random Gaussian noise with a 5\% amplitude.
We create six datasets in which we input a deformation $\adata\in\{0.25,0.5,1,1.5,2,2.5\}$. In each dataset, we make 10 mock bubble stacks (for which the noise is generated randomly) to simulate a larger volume. A mock bubble stack is shown for each set in Fig. \ref{fig:toymodel_maps}, where in each panel, a different universe is shown with the different $\adata$ values.

After computing the histogram ratios, which are shown in Fig. \ref{fig:toymodel_PDFangles}, we are qualitatively able to determine the deformation $\adata$ of each mock bubble stack. Again, each panel represents a different universe with a deformation value $\adata\in\{0.5,1,2\}$. The colors represent different $\alpha$ deformations tested in our model. Taking the panel where the truth is the same as the fiducial ($\adata=1$, middle panel) as an example, we can see that $H_{N}/H_{\overline{N}_{\alpha}}$ is flat when $\alpha=1.00$ in our model. This means that the modeled bubble stack $\bar{N}_1$ shows no particular direction in which it would be different from the original bubble stack $N$. We can thus qualitatively say that the deformation parameter $\adata$ is near 1.00 in this universe, which corresponds to the expected $\adata=1$ value. For the other universes (other panels), we can also see that the expected $\adata$ can be qualitatively inferred.

To better quantify and constrain the deformation of the stacks, we can look at the $\chi^2$ statistic, shown in Fig. \ref{fig:toymodel_chi2}. Each color represents a different universe with a deformation value $\adata\in\{0.25,0.5,1,1.5,2,2.5\}$. In this toy model case, the covariance matrix of Eq. \ref{eq:chi2cov} is diagonal, meaning that all $\theta$ bins are independent of each other. For each dataset, its diagonal elements are thus the standard deviation of the histogram ratio obtained from the 10 bubble stacks.
For each Universe, we can see that the $\chi^2$ is minimized at an $\amin$ value that is close to that of the input cosmology.

This toy model therefore helps us to illustrate our AP test, also showing that it can give us constraints on the deformation $\adata$ present in a mock bubble stack. The constraints we find for this case are very precise, with errors at the percent level. This is of course related to the level of noise we added in this example, and decreasing the signal-to-noise ratio makes the error bars larger.
In the following, we describe what constraints we can put on $\adata$ for bubbles simulated in a more realistic manner.

\section{From 21-cm to Bubble Stacks} \label{sec:unet_bubbles}

A promising way to observe reionization bubbles is the 21-cm signal, which traces neutral hydrogen, making it therefore directly dependent on the ionization state of the IGM. The observable at hand is the 21-cm brightness temperature \cite{Furlanetto2006,Pritchard2012}:
\begin{equation}
    \dTb \approx T_0(z) \xHI (1+\delta) \left(1-\frac{\Tcmb}{T_s}\right),
\label{eq:dtb}
\end{equation}
which measures the deviation (absorption or emission) from the CMB sourced by these atoms. It depends on the position and redshift at which it is observed, as well as on the neutral ionization fraction  $\xHI$, the gas density $\delta$, as well as the CMB and spin temperatures ($\Tcmb$ and $T_s$, respectively).
Here $T_0$ is a normalization factor, defined as
\begin{equation}
    T_0(z) = 34 \text{ mK} \times \left(\frac{1+z}{16}\right)^{1/2} \left(\frac{\Ob h^2}{0.022}\right) \left(\frac{\Om h^2}{0.14}\right)^{-1/2},
\label{eq:T0}
\end{equation}
which depends on the cosmology through the reduced Hubble parameter $h$, and the baryon and matter densities $\Ob$ and $\Om$.

Because the 21-cm signal is emitted by neutral hydrogen atoms, it obviously depends on whether or not the gas is ionized through $\xHI$: there is no signal at the places where all the hydrogen atoms are ionized. These no-signal regions correspond directly to the ionized bubbles that we are interested in. The 21-cm signal can therefore in principle allow us to tomographically map the ionized bubbles on the sky as a function of redshift.

Current observations are setting limits on the 21-cm power spectrum, and are inching towards a detection of this observable \citep{Mertens2020,Trott2020,HERA2023}. However, a next-generation instrument like the SKA is expected to obtain 21-cm light-cones (i.e., 3D maps), from which to extract ionized bubbles.
There will, of course, be instrumental noise in these observations, and the signal will be affected by astrophysical foreground contamination. 
This will be the main obstacle in recovering the ionized bubbles, as we explore in this section.

\subsection{Mock 21-cm observations} \label{sec:unet_mockobs}

\begin{figure*}
\centering
\includegraphics[width=1\textwidth]{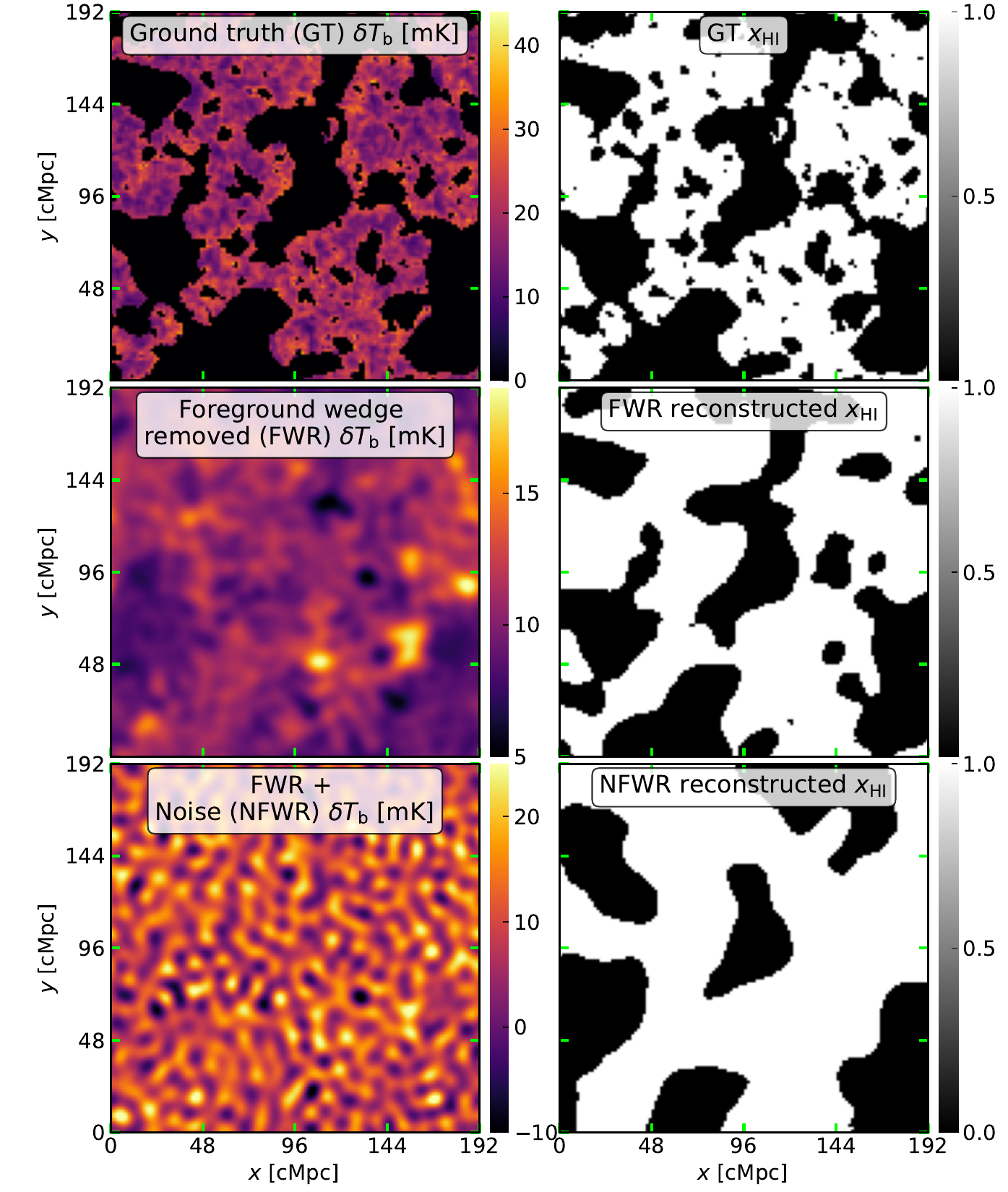}
\caption{Illustration of how we reconstruct $x_{\rm HI}$ ({\bf right}) from 21-cm maps $\dTb$ ({\bf left}) in the presence of noise and foregrounds, where we show slices of our simulations on the plane of the sky at $z=7.5$. The {\bf top} panels correspond to the ground truth ($\cmfast$) maps, which can be thought of as the training set. The {\bf middle} panels show the wedge-removed $\dTb$ map on the left and the U-net recovered $\xHI$ map on the right (both belonging to the FWR dataset). The {\bf bottom} panels  have noise in addition to a foreground-wedge-removed $\dTb$ (NFWR), so the U-net prediction for the $\xHI$ map on the right is less precise. 
While by eye one cannot obviously find the ionized regions within the foreground wedge-removed $\dTb$ maps of the middle and bottom left panels, the U-net exploits non-Gaussian correlations to reconstruct the $x_{\rm HI}$ maps in the middle and bottom right panels. The quality of the reconstruction worsens due to noise in the bottom panel, especially at small scales, which we account for in our analysis.
\label{fig:unet_maps}}
\end{figure*}

In order to create 3D mock 21-cm boxes, we start by generating the 21-cm signal using the semi-analytical code $\cmfast$ \citep{Mesinger2011,Muray2020} following \citet{Kennedy2024}. We use the default astrophysical and cosmological~ \cite{PlanckCollab2018} parameters of $\cmfast$ to obtain many 3D 21-cm coeval boxes with different random seeds. The resolution of these simulations is $\Delta x=1.5$ $\cMpc$, and the boxes have $128^3$ voxels (which corresponds to a $192^3$ $\cMpc^3$ volume). In this first work, we will focus on 21-cm boxes taken at redshift $z=7.5$, and comment on other redshifts in Appendix \ref{app:z_lightcone_effects}. We note that redshift space distortions (RSD) are by default taken into account in our simulation boxes, so any induced ``squashing'' of bubbles along the line of sight does not impact our results within the error-bars quoted.
The $\dTb$ maps generated this way will be later called ground truth (GT), as they will be used to create input maps for a neural network detailed in the following section. An example of such a map is shown in the top left panel of Fig. \ref{fig:unet_maps}. The top right panel is the corresponding neutral fraction map, where white regions are neutral and black regions are ionized. By comparing these last regions to the 21-cm map on the left, we can see that the $\dTb=0$ regions indeed correspond to $\xHI=0$. 

The observed 21-cm signal will be highly affected by astrophysical foreground contamination. Indeed, Galactic and extragalactic foregrounds will be three to four orders of magnitude stronger than the 21-cm signal \citep{Bernardi2009,Bernardi2010}. 
Luckily, these foregrounds are contained to a region in Fourier space named the ``foreground wedge'' that affects large-scale line-of-sight wavenumbers ($k_{||}$) more than their angular ($k_{\perp}$) counterparts~\citep{Liu:2019awk}.
In principle we could restrict ourselves to the complementary foreground-free region, called the ``EoR window'', as is often done for 21-cm cosmology studies (e.g.,~\cite{HERA2023}). 
We see that the foreground removal procedure has a great impact on the 21-cm maps (middle-left panel) as it blurs and distorts the signal so it becomes difficult to distinguish the ionized regions (the darkest regions do not really correspond to bubbles anymore).
The ionized bubbles are structures in configuration space, as we can see in Fig.~\ref{fig:unet_maps}, so to find them we need all the Fourier modes, including those missing because of the wedge. Fortunately, the non-Gaussianity of the 21-cm signal allows us to recover information on the modes inside the wedge with only outside-wedge data~\cite{Li2019,Makinen2021,GH2021,Bianco2024,Kennedy2024,Sabti2024,Bianco2024b,Beardsley:2014bea}, as we will explore in the next subsection.

In addition to foregrounds, we want to make the boxes more realistic by adding instrumental noise corresponding to the SKA radio telescope. To do that, we use the $\toolscm$ Python package \citep{Giri2020} setting an integration time $t_{\text{int}}=10$ s, a daily observation time $t_{\text{obs,day}}=6$ hrs, and a total observation time $t_{\text{obs,tot}}=2000$ hrs, following the procedure in \citet{Kennedy2024} (which can be adjusted with a higher observation time if the observed 21-cm fluctuations were smaller than our fiducial). 
The resulting noisy and foreground-removed (NFWR) map can be seen in the bottom-left panel of Fig.~\ref{fig:unet_maps}, where the 21-cm signal is dominated by noise and bubble structures are no longer visible by eye.

\subsection{Wedge-recovered ionization maps} \label{sec:unet_xHI}

Let us now describe how we recover ionized regions from foreground-removed 21-cm maps through a deep-learning approach that reconstructs the modes lost in the anisotropic foreground wedge by exploiting the non-Gaussianity inherent to the field.
This method is based on a neural network first developed by \citet{GH2021}, and then improved by \citet{Kennedy2024}, which is the version that we use in this work. This network is a U-net that does not need any knowledge on the foregrounds or any explicit knowledge on the fact that ionized regions distribute as bubbles. Its input is a foreground wedge-removed $\dTb$ box (with or without noise, where the wedge-removal procedure consists of zeroing out the modes that are in the foreground wedge), normalized between 0 and 1. The U-net is trained to reproduce a binarized $\xHI$ map (which we refer to as the prediction or PRED in this paper) that should thus contain information from both inside and outside of the foreground wedge.
\citet{Kennedy2024} showed that this U-net is able to reconstruct $\xHI$ boxes from foreground wedge-removed $\dTb$ boxes during reionization with a good accuracy, as we corroborate in the middle and bottom right panels of Fig.~\ref{fig:unet_maps}. Without instrumental noise, we see that the neural network can indeed recover the large-scale morphology of the reionization map (middle-right panel), though not the small-scale features. When noise is added, it dominates the signal and the recovered ionized regions (bottom-right panel) are even coarser, in agreement with the findings of
\citet{Kennedy2024} at higher redshifts (where noise is higher).
This reconstruction will then fail at small scales, or when noise is too high, but suffices for the purposes of this work.

We use \citet{Kennedy2024}'s U-net to predict binarized $\xHI$ boxes at $z=7.5$ from the two wedge-removed $\dTb$ datasets presented in Sec \ref{sec:unet_mockobs}, one without  (called FWR1) and one with noise (called NFWR1), in all cases with standard cosmology (and thus no deformation, $\adata=1$). 
We also train the U-net with a fixed set of astrophysical parameters (the default ones of $\cmfast$). As it is, the neural network is therefore potentially model dependent, and there are two ways to circumvent such a dependence. One possibility is that by the time SKA provides 21-cm light cones, there may already be constraints on astrophysical parameters from 21-cm power spectrum measurements. We could, thus, retrain our neural network within the error bars of such a measurement and re-do the analysis of this paper, as suggested in \citet{Sabti2024}. A second option is to train the neural network on 21-cm signals obtained from a series of different reionization models (within the regimes favored by observations). This way, the neural network would not learn features in the 21-cm signal of a specific model, and would rather allow us to obtain a more generic prediction for the $\xHI$ map given $\dTb$ while being agnostic of the underlying astrophysical model \citep{Zhou2022}. We explore this option in Appendix \ref{app:varying_params}, where we find that a model-agnostic U-net can still recover the $\xHI$ maps well enough to perform our AP test.

We want to additionally test cases where $\adata\neq1$ to ensure that the deformation does not affect the bubble recovery and that we can recover the input cosmology if there was deformation. Thus, we also deform our $\dTb$ boxes (before adding noise or wedge-filtering) along the line of sight ($z$ axis) with two different deformation levels: $\adata=0.8$ and $\adata=0.9$, corresponding respectively to 20\% and 10\% contraction. Our contraction method consists of averaging slices of a larger box along the $z$ axis. For example, for a $\adata=0.8$ contraction, we use $256^3$ boxes (while keeping a $1.5^3$ $\cMpc^3$ volume resolution), and along the line of sight ($z$ direction) and for every ten successive slices, we create eight slices that are the average of side-by-side slices. 
We then give the two deformed datasets (deformed FWR and NFWR 21-cm boxes) to the U-net that is trained on the non-deformed boxes, and obtain four more datasets of contracted reconstructed $\xHI$ boxes (FWR.8, FWR.9, NFWR.8 and NFWR.9, the notation numbers corresponding to the $\adata$ deformation). 
We made the choice to run the deformed mock $\dTb$ datasets through a U-net trained in the fiducial cosmology (non-deformed $\dTb$ boxes) because, as we will not know the true cosmology, we want to make sure that the U-net preserves deformations and that our AP test gives proper constraints with $\xHI$ predictions obtained from the fiducial cosmology.

\subsection{Constructing the bubble stack} \label{sec:unet_BS}

\begin{figure*}
\centering
\includegraphics[width=1\textwidth]{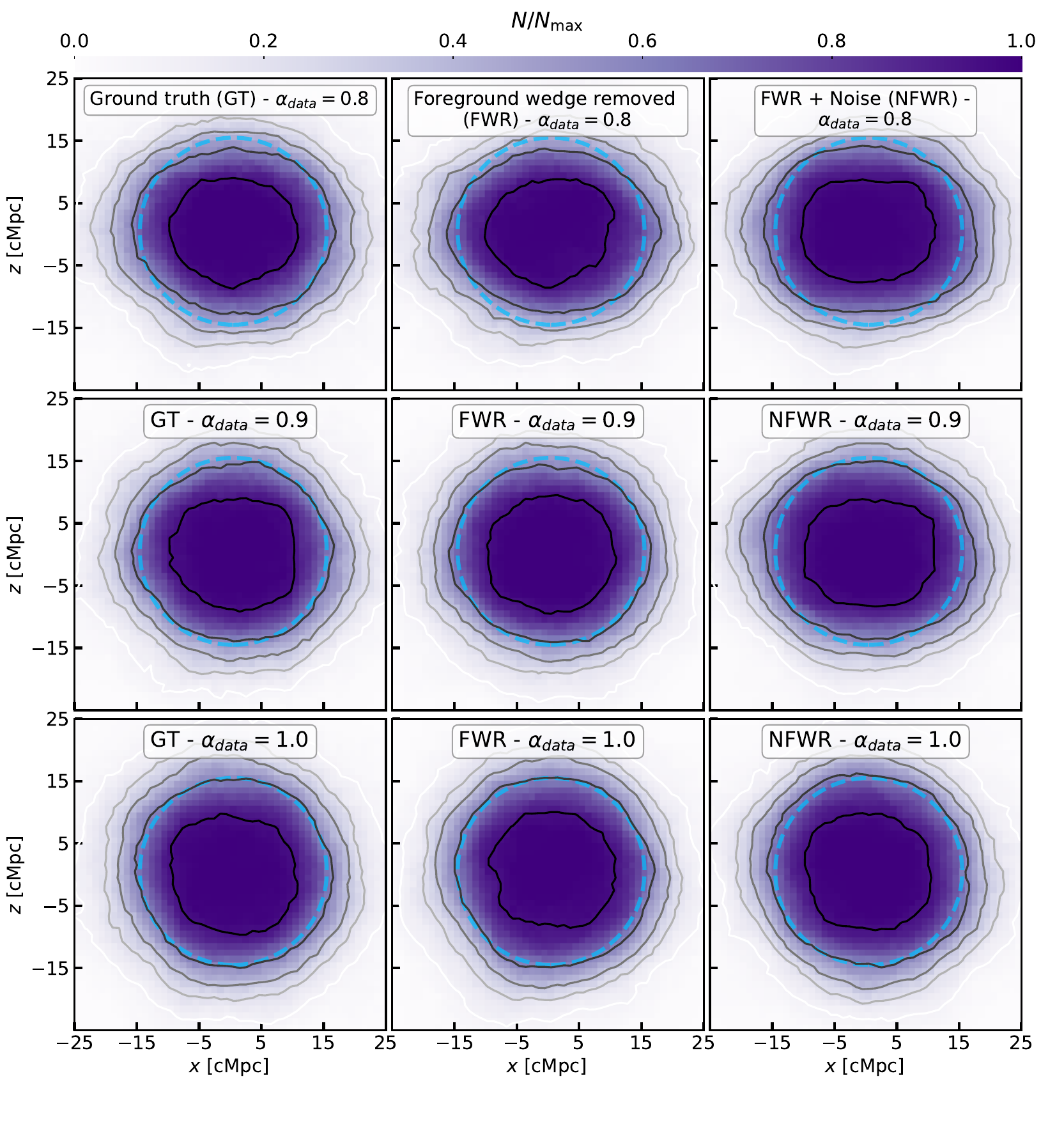}
\caption{Bubble stacks extracted from the 21-cm signal, in all cases assuming a volume close to 1 cGpc$^3$ as expected from SKA observations at $z = 7.5$. 
The left column shows the ground truth (GT, directly extracted from 21-cm simulations). 
The middle and right columns show bubble stacks from the U-net reconstructed $\xHI$ boxes, with only foreground removal (FWR, middle), and also with noise (NFWR, right). 
In all cases we keep only bubbles with a radius $\rbubbles>25$ $\cMpc$ to avoid artifacts with smaller bubbles when adding noise. 
For ease of visualization, here we normalize each bubble stack to its maximum value. In each row, we show a different universe in which the deformation $\adata$ is 0.8, 0.9 or 1 from top to bottom. Isocontours of the number of bubbles are shown in gray scale. We have added the same blue dashed circle at $\rbubbles=15$ cMpc in every panel to help guide the eye. In the case $\adata=1$, we are in the Universe with the fiducial cosmology, and the bubble stacks are spherical, whereas in the rest we can observe contraction even with noise and foreground removal.
\label{fig:unet_bubblestacks}}
\end{figure*}

In order to create bubble stacks for each dataset presented above, we need to extract the ionized bubbles, both from the $\dTb$ maps and from the U-net reconstructed $\xHI$ boxes. 
We do so through a watershed algorithm as presented in \citet{Lin2016} (and encourage the interested reader to visit this paper for further details on how this algorithm works, as we will only review it briefly here). 
The basic idea behind the watershed algorithm can be pictured with the following 2D analogy: imagine a hilly landscape, in which one pours water until all the valleys are filled. The water will then be separated in different basins, which are an analogy to our ionized bubbles here. 
The watershed code will hence associate to each pixel $i$ of a binarized field (that can be 2D or 3D) the distance $-d_i$ to the closest pixel in which the binary value is different (we use the opposite $-d_i$ of the Euclidean distances $d_i$ to keep the standard watershed terminology). The minima of this distance field are the centers of the basins (or the bubbles here), and its isocontours (also called ``watershed lines'') correspond to the edges of the basins. 
Moreover, one can tune one parameter in this code that is called the ``h-minimum transform''  $\hws$ to prevent over-segmenting the field, as all the local minima of the field could otherwise have their own basin \citep{Lin2016}. This $\hws$ parameter (in units of pixels) is a distance threshold that will smooth out the shallow local minima that are potentially due to noise in the data. In this paper, we use $\hws=0.6$, so that all the local minima (or bubbles) are retrieved while the box is not over-segmented. \citet{Lin2016} shows that having a too high $\hws$ value would bias the bubble size distribution (BSD) towards larger bubbles. They also establish that there is a range of values containing $\hws=0.6$ in which the segmentation in bubbles is fairly stable and the corresponding BSDs are visually close to each other. Within this range, they find good agreement in BSDs with the physically motivated mean free path method. 

Furthermore, we want to highlight that the default watershed algorithm is excellent at detecting spherical bubbles \citep{Lin2016}, so much so that in deformed boxes it has a tendency to find more spherical bubbles than it should. We have therefore customized the \citet{Lin2016} watershed code so it is able to find bubbles having the proper shape when we change their topology by deforming the boxes. This enhancement is discussed in Appendix \ref{app:watershed_appendix}.
We thus use this version of the watershed algorithm to extract the bubbles from each U-net-reconstructed $\xHI$ (or PRED) box for every input $\alpha$ (FWR.8, FWR.9, FWR1, NFWR.8, NFWR.9 and NFWR1, including different initial seeds).

The U-net recovered (PRED) boxes are already binarized. The GT 21 cm boxes are binarized as follows: a cell is ionized if $\dTb=0$ and neutral otherwise. 
From each GT box, and each corresponding PRED boxes of every dataset, we extract the ionized bubbles by running the watershed algorithm, and we keep all the bubbles that have a radius $\rbubbles>25$ $\cMpc$. We make the choice of keeping only the largest bubbles because of the observational noise: they are more robust than their smaller counterparts \citep{Kennedy2024}.
The geometrical centers of the bubbles are computed, and all the bubbles are centered on the same voxel, so that we can stack them to form the bubble stacks $N$. A stack is composed of all the bubbles (with $\rbubbles>25$ $\cMpc$) present in 25 GT or PRED boxes respectively to simulate the future sky surface observed by the SKA (which is close to 1 $\cGpcSquare$). 
In addition, we have tested several number of bubble stacks to compute the covariance matrix $\mathbf{C}$ of Eq. \ref{eq:chi2cov}, and concluded that with 400 bubble stacks, $\mathbf{C}$ has converged \citep{Taylor2013}. We therefore create 400 GT bubble stacks, as well as 400 PRED bubble stacks for every dataset. 

An example of GT and FWR and NFWR PRED bubble stacks is shown in the different columns of Fig. \ref{fig:unet_bubblestacks}. Each row represents a different universe, in which we deformed the GT boxes with $\adata=$ 0.8, 0.9 and 1 from top to bottom. 
We have drawn a blue circle on the bubble stacks to help the eye see the deformation or sphericity of our bubble stacks. Visually, the $\adata=0.8$ and $\adata=0.9$ bubble stacks (first and second rows) are indeed contracted, and the $\adata=1$ (last row) stacks are rather spherical. Also, by eye the different deformations shown here seem to be the same whether we look at the GT maps or the predicted FWR and NFWR maps.
The U-net therefore preserves the contraction applied to the input boxes and can reliably reproduce the observed bubble shape for us to perform the AP test.

\section{AP test on bubble stacks} \label{sec:results}

In this Section, we present the results of our AP test on the data sets described in Sec. \ref{sec:unet_bubbles}. We start by checking if we are able to recover the input deformation $\adata$ applied on each set of bubble stacks, and forecast constraints on $\Da H$ at reionization redshifts.

\subsection{Resulting constraints on $\adata$} \label{sec:results_alpha}

\begin{figure}[!]
\centering
\includegraphics[width=0.49\textwidth]{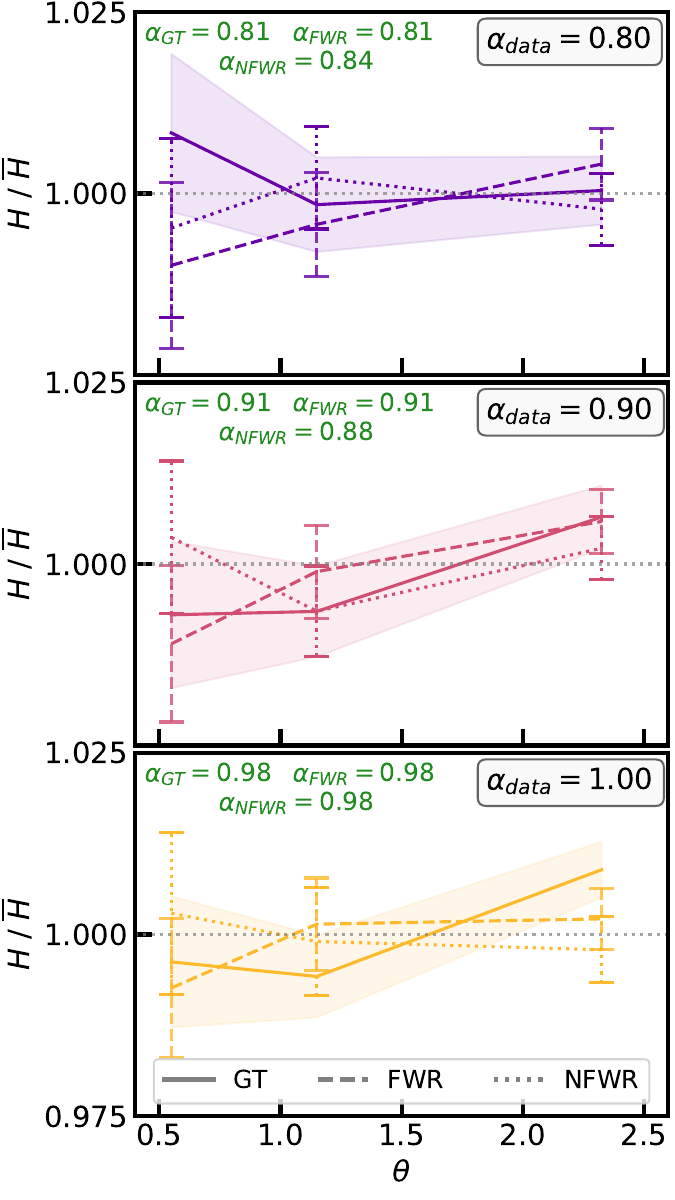}
\caption{Ratio of histograms (between that of a bubble stack $H$ and its closest spherical counterpart $\overline H$) for a bubble stack from the ground truth $\dTb$ (GT, full line), from the foreground-wedge-removed U-net reconstructed $\xHI$ maps (FWR PRED, dashed lines), and the same but when adding noise (NFWR PRED, dotted  line). 
Unity in $H/\bar{H} (\equiv H_{N}/H_{\overline{N}_{\alpha}}$) means we have found the correct cosmology in the form of a deformation $\alpha = \alpha_{\rm data}$.
Each panel represents a different universe in which the input deformation $\adata$ is 0.8, 0.9 or 1 from top to bottom. For each dataset, we show the best-fit line as inferred by the $\chi^2$, with the $\alpha$ value specified in green. The shaded area and error bars represent the standard deviation calculated from the covariance matrix of each data set. In each panel, we can see that the best-fit histogram is roughly unity, meaning that we are close to expectations with these $\alpha$ values.
\label{fig:unet_PDFangles}}
\end{figure}

\begin{figure}[!]
\centering
\includegraphics[width=0.5\textwidth]{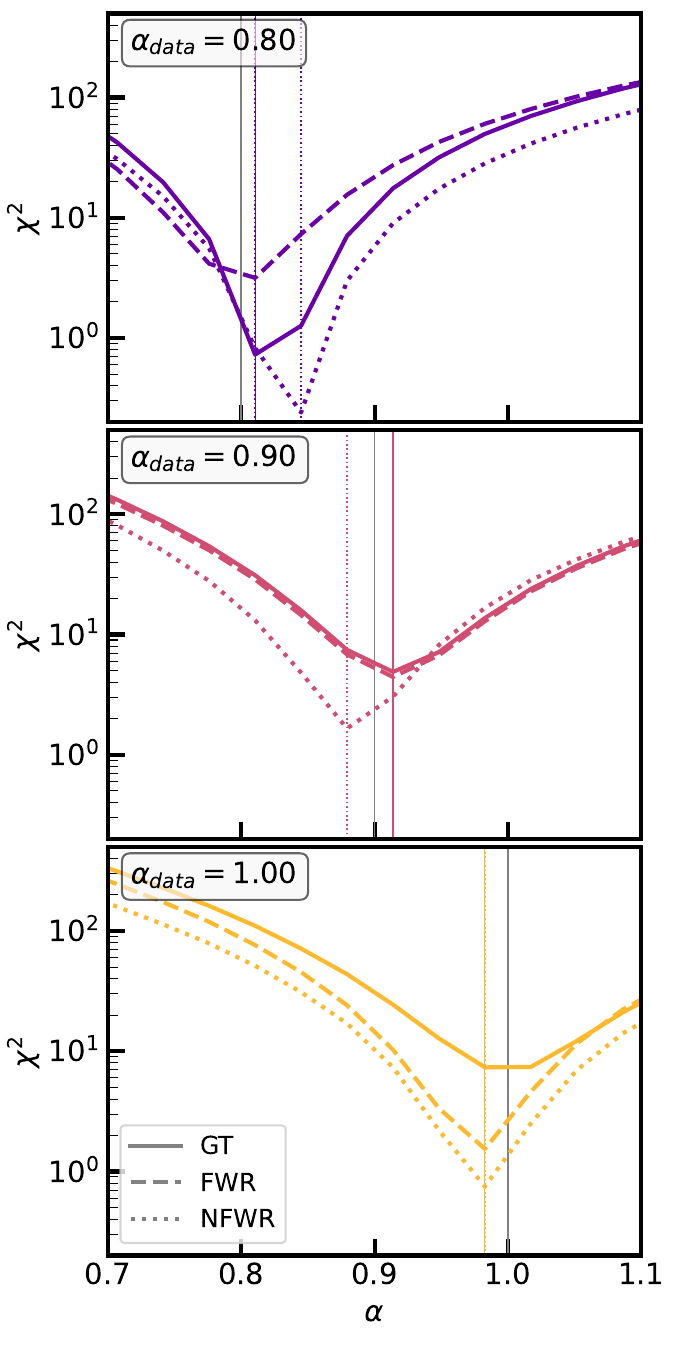}
\caption{$\chi^2$ test of each of our bubble stacks (GT in full lines, FWR in dashed lines, and  NFWR in dotted lines) against $\alpha$. Each panel represents a different universe with a given $\adata\in\{0.8, 0.9, 1\}$. The $\adata$ deformation can be inferred by looking at the value of $\alpha$ at which the $\chi^2$ is minimal, and we compute error bars by taking the errors at 68\% CL from the $\chi^2$. For each case tested here, the deformation is correctly inferred within the error bars, and the forecasted constraints are summarized in Table \ref{tab:unet_datasets}. The gray vertical line is the $\alpha=\adata$ line, and the colored ones correspond to the $\alpha$ at which the $\chi^2$ is minimized.
\label{fig:unet_chi2}}
\end{figure}

For each bubble stack extracted from the 21-cm signal (directly from the GT $\dTb$ boxes or via the U-net from the FWR and NFWR PRED $\xHI$ boxes), we compute the histogram ratio of Eq. \ref{eq:finding_adata}, that is, the ratio between the histogram of the map and its spherically symmetrized version.
These are shown in Fig. \ref{fig:unet_PDFangles}
for different Universes with input deformations $\adata=$ 0.8, 0.9 and 1.
We first want to highlight that we want a $\theta$ resolution that enables us to resolve the departures from $H/\overline{H}=1$. However, as the $\theta$ bins are correlated, the covariance matrix $\mathbf{C}$ is not diagonal and is thus harder to compute, and requires more bins and stacks \citep{Taylor2013}. We therefore adopt the compromise of having three angular bins and 400 stacks, and have tested convergence of the covariance matrix elements.

Now, looking at every histogram ratio in Fig.~\ref{fig:unet_PDFangles}, we can see that they roughly agree with unity within the error bars for the best-fit $\alpha$ values shown for each universe, as expected if we have correctly found the deformation level (i.e., $\Da H$). The error bars shown here are from the diagonal elements of the covariance matrix, and are thus computed from the 400 bubble stacks for each dataset, corresponding to statistical errors on $H/\bar{H}$ (though the histogram ratios are computed from a unique bubble stack, representing the entire SKA field of view). In all cases, including the GT, $H/\bar{H}$ is not perfectly flat, and some error bars do not reach 1, indicating possible systematics, as we will discuss near the end of this section. 

We compute the $\chi^2$ using Eq. \ref{eq:chi2cov} for each of these datasets, and show them in Fig. \ref{fig:unet_chi2}. The minima of each curve are shown in Table \ref{tab:unet_datasets} along with the forecasted errors at 68\% confidence level (CL). Here again, the $\chi^2$ curves come from a unique bubble stack, and we note that the binning in $\alpha$ can slightly impact the shape of the $\chi^2$.
Looking at each dataset, the $\amin$ deformations recovered agree with the input within the error bars, except for the FWR $\adata=1$ case where we get an $\amin=0.98^{+0.01}_{-0.02}$ at 68\% CL, which is marginally lower than expected, though not significantly so. 
In all cases, we forecast error bars at the percent level, showcasing the power of this AP test to extract cosmology during reionization.

Before interpreting our forecasted constraints, let us discuss a few limitations. 
We have only used co-evaluated (i.e., same $z$) boxes at $z=7.5$ and stacks with $\rbubbles>25$ $\cMpc$ bubbles. 
For the former, we show in Appendix \ref{app:z_lightcone_effects}, the results at different redshifts, as well as using light cones (where higher redshifts along the line-of-sight represent higher comoving distances). We find that using light cones does not alter the result of our AP test. We also show that our AP test can be applied on bubbles observed at redshifts 6 to 11 (which spans all the EoR with our reionization model).
The cutoff of $\rbubbles>25$ $\cMpc$ bubbles is motivated by the fact that when the data are noisy, the U-net struggles to recover properly the smaller bubbles, so small-scale structures are driven by noise and are thus roughly spherical \citep{Kennedy2024}. As long as we are able to extract enough bubbles statistically, focusing on the larger bubbles makes this probe more robust to noise.
Finally, we note that all of our uncertainties are computed by only taking into account statistical errors, which means that we could lack systematic errors. 
These include possible issues with the watershed algorithm, leakage of foregrounds above the wedge~\citep{Cunnington:2020njn}, or exotic noise properties. 
As an example, when using bubbles with $\rbubbles= 5-25$ $\cMpc$ we find that the NFWR $\adata=1$ case produces a systematic offset of a few percent in $\alpha$ on the $\chi^2$ as the watershed algorithm recovers bubbles that are less spherical than they are in the noisy maps.
This could be mitigated through detailed simulations to account for these systematics. 
Nevertheless, the tests described above (and in Appendix \ref{app:z_lightcone_effects}) show that our test is robust to the 2\% level, given our current understanding of 21-cm maps and foregrounds, which ought to be revisited once the SKA is online.

\begin{table}[!]
    \centering
    \normalsize
    \setlength{\tabcolsep}{0.25cm}
    \begin{tabular}{l|ccc}
    \hline \hline \\[-0.25cm]
    $\adata$ & \textbf{0.8} & \textbf{0.9} & \textbf{1} \\ [0.15cm]
    \hline \\[-0.25cm]
    GT & $0.81^{+0.04}_{-0.01}$ & $0.91^{+0.02}_{-0.01}$ & $0.98^{+0.04}_{-0.01}$ \\ [0.25cm]
    FWR & $0.81^{+0.01}_{-0.03}$ & $0.91^{+0.02}_{-0.01}$ & $0.98^{+0.01}_{-0.02}$ \\ [0.25cm]
    NFWR & $0.84^{+0.01}_{-0.04}$ & $0.88^{+0.03}_{-0.01}$ & $0.98^{+0.02}_{-0.02}$ \\ [0.15cm]
    \hline
    \end{tabular}
    \caption{Constraints inferred for the AP parameter of the data $\adata=\Da H/(\Da H)'$ with errors at 68\% CL. GT stands for ground truth data and represents the bubble stacks made from simulated $\dTb$ maps. FWR stands for Foreground Wedge-Removed and NFWR for Noisy Foreground Wedge-removed, and both stand for bubble stacks made from U-net reconstructed $\xHI$ maps from foreground-removed $\dTb$ maps, either without or with instrumental noise. In all cases, except the FWR $\adata=1$ case that is slightly lower than expected, we recover the input $\alpha_{\rm data}$ within 68\% confidence, and for the observationally realistic NFWR we expect $\sim 2\%$ forecasted errors.}
    \label{tab:unet_datasets}
\end{table}

\subsection{$\Da H$ constraint forecasts} \label{sec:results_interpretation}

\begin{figure*}
\centering
\includegraphics[width=1\textwidth]{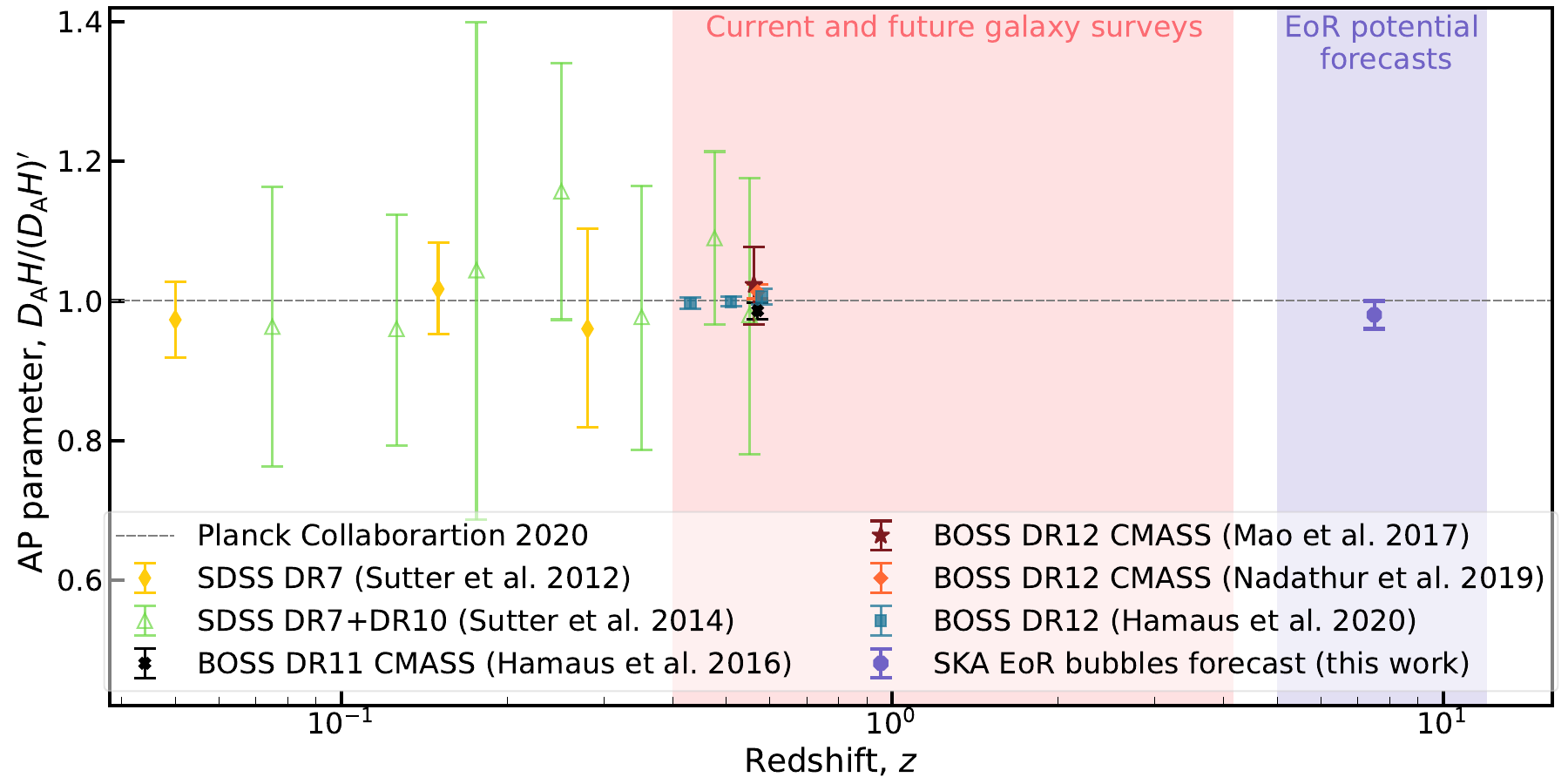}
\caption{Current constraints and our forecast (both with 68\% CL) for $\Da H$ compared to the fiducial $(\Da H)'$ from \citet{PlanckCollab2018}. The data points at redshifts $z<1$ are current constraints obtained from galaxy surveys \citep{Sutter2012,Sutter2014,Hamaus2016,Mao2017,Nadathur2019,Hamaus2020}. The red area shows the redshift range in which ongoing galaxy surveys, like DESI (Dark Energy Spectroscopic Instrument; \cite{DESIColla2024}), HETDEX (Hobby-Eberly Telescope Dark Energy Survey; \cite{Gebhardt2021}) or PFS (Subaru Prime Focus Spectrograph; \cite{Okumura2022}), can constrain the AP parameter. The purple point at $z=7.5$ is our forecast using ionized bubbles extracted from observations of the 21-cm by the SKA. The purple shaded area shows where we expect to be able measure $\Da H$ with our AP test.
\label{fig:DaH_constraints}
}
\end{figure*}

Table \ref{tab:unet_datasets} summarizes the $\Da \times H$ forecasts we obtain in this work.
Each row in the table considers different cases of observational noise: GT is the case without any noise, FWR is contaminated by the foreground wedge, and NFWR additionally has instrumental noise. 
Each column tests a different input cosmology.
In all cases we are able to retrieve the input deformation (and thus $\Da H$), which confirms that our AP test also works for observations away from our fiducial cosmology.
Focusing on the NFWR mock observations with a \citet{PlanckCollab2018} cosmology, as this is the closest to the analysis we will perform with SKA or other 21-cm data, our forecast can be recast as
\begin{align}
    \alpha &= \frac{\Da H}{(\Da H)'} = 0.98^{+0.02}_{-0.02} \quad\text{or}\quad \nonumber\\ \Da H/c&=3.24^{+0.07}_{-0.07}, \quad\text{at} \quad z=7.5
\label{eq:forecasts}
\end{align}
at 68\% CL, where $(\Da H)'/c=3.30$ in a fixed \citet{PlanckCollab2018} $\Lambda$CDM cosmology at $z=7.5$, and $c$ is the speed of light. 
This demonstrates that we can put constraints on this product of cosmological parameters with the volume expected in SKA observations. We want to stress again that we can only measure the product of $\Da$ and $H$, and not on each of them individually. 

We summarize in Fig. \ref{fig:DaH_constraints} the existing constraints on $\Da H$ with respect to \citet{PlanckCollab2018} that have been obtained thanks to AP tests with cosmic voids (see \citet{Hamaus2020}). The red area shows the redshift range in which the DESI \citep{DESIColla2024}, HETDEX \citep{Gebhardt2021}, and PFS \citep{Okumura2022} surveys are able to put constraints using BAO measurements (i.e., calibrated standard rulers). Currently, we do not have constraints on this product of parameters at reionization redshifts (though there are recent forecasts for Cosmic-Dawn and EoR redshifts using standard rulers~\cite{Munoz2019,Fronenberg2024}).
The purple point at redshift $z=7.5$ corresponds to our AP bubble test forecast. We can also see that the statistical errors we obtain in this work are comparable to the errors on the low-$z$ data points.

As detailed before, we focused on $z= 7.5$ in this work, but our AP test is feasible during all reionization as long as we are able to distinguish enough ionized bubbles in our data (at least a few hundred, see Appendix \ref{app:z_lightcone_effects}). The purple box in Fig. \ref{fig:DaH_constraints} represents the possible redshifts covered by this technique. 
The higher $z$ limit of this box is set by the detectability of bubbles: at the beginning of reionization, the bubbles are too small to be resolved even with next-generation 21-cm experiments. The lower $z$ limit is set by the percolation of bubbles, as eventually we will not be able to distinguish individual bubbles at the end of reionization.
The exact location of the high- and low-$z$ limits depends on the timing of reionization~\citep{Robertson:2015uda,Munoz:2024fas}, so they are only approximate.

\section{Conclusions} \label{sec:conclusion}

In this work, we have presented a new AP test to measure the product of cosmological parameters $\Da H$, with $\Da$ being the comoving angular diameter distance and $H$ the Hubble rate, using ionized bubbles during reionization (at $z\sim 5-12$). Because the ionizing fronts propagate isotropically on average, a statistically large stack of bubbles will be spherical, even if any individual bubble is not. A bubble stack is therefore a standard sphere, which we use in an AP test. This kind of test is purely geometric, and allows us to constrain the product $\Da H$ independently of astrophysics and cosmology.

We have tested that this AP test is robust and unbiased. 
The first challenge is observing bubbles during reionization. We used the 21-cm signal as a tracer of $x_{\rm HII}$ bubbles as it depends directly on the neutral hydrogen fraction. The SKA will produce tomographic 21-cm maps of the sky, but the cosmic signal will be diluted in astrophysical foregrounds and instrumental noise, making it complicated to extract  ionized bubbles. We circumvent this issue by using a neural network (from \citet{Kennedy2024}) to reconstruct the ionized regions from the noisy foreground-removed 21-cm signal.
We extract bubbles with a radius $\rbubbles>25$ $\cMpc$ from the reconstructed $\xHI$ maps, as we find that large bubbles are more robust than smaller ones, and take a volume corresponding to the whole SKA field of view. We then run a sphericity test on the bubble stacks, in which we compare each bubble stack to its spherically averaged counterpart.
With our bubble stacks, we find that we will be able to measure $\Da H$ to $\sim 2\%$ precision.
We have additionally tested that we can recover 
non-standard cosmologies by contracting the input boxes along the line of sight by 10\% and 20\%, 
and have considered realistic foreground avoidance by removing the foreground wedge from the the 21-cm signal, both with and without SKA instrumental noise. 
In almost all cases we can recover the correct input value of $\adata$ within 68\% CL. Only one case, the non-deformed foreground wedge-removed bubble stacks, showed a slight deviation, although not significant.

Another challenge is whether the test would hold for a neural network not trained on the correct astrophysics.
We perform a limited test for this in Appendix~\ref{app:varying_params}, and find that we can still obtain proper constraints while being robust to parameter variation over a wide range of reionization scenarios (all within the $\cmfast$ framework, however).
We also checked our AP test at other redshifts than 7.5, as well as with light-cones instead of coeval boxes, in Appendix \ref{app:z_lightcone_effects}, and concluded that it also works in these cases. 
While there may remain systematics below the few \% level, our tests suggest that our AP test will be able to put constraints on $\Da H$ through reionization with 21-cm light cones from the SKA.
In the meantime, one could hope to observe ionized bubbles using different probes (e.g. with Ly$\alpha$ observations, see \cite{Mukherjee2024,Lu2024}). These kinds of observables have very different observational and theoretical systematics, so they are beyond the scope of this work, but first studies show great promise. 

In summary, we have shown that stacks of reionization bubbles act as standard spheres, allowing us to perform an AP test at high redshifts. 
This work is a proof of concept, and we find that as long as we observe enough (at least a few hundred) ionized bubbles, we can constrain $\Da H$ during reionization to $\sim 2 \%$ precision. 
Given the cost of traditional galaxy surveys at these redshifts, this probe has the potential to measure cosmic geometry between us and the CMB, filling the missing chapters in our cosmic history.

\begin{acknowledgments}

We are thankful to Dominique Aubert, Neal Dalal, Hannah Fronenberg, Joohyun Lee, Jacob Kennedy, Debanjan Sarkar, and Yonatan Sklansky for discussions and comments on this manuscript.
This work was supported at UT Austin by NSF Grants AST-2307354 and AST-2408637, and by the NSF-Simons AI Institute for Cosmic Origins.
AL and FDB acknowledge support from the Trottier Space Institute, a Natural Science and Engineering Research Council of Canada (NSERC) Discovery Grant, an NSERC-Fonds de recherche du Québec – Nature et technologies (FRQNT) NOVA grant, and the William Dawson Scholarship at McGill. FDB was additionally supported by an NSERC Undergraduate Student Research Award (USRA) as well as an FRQNT NSERC USRA supplement.
\end{acknowledgments}

\bibliography{biblio}

\appendix

\newpage
\section{Redshift and light-cone effects} \label{app:z_lightcone_effects}

Through the main text we focused on results for 3D 21-cm signal boxes at a fixed (co-eval) redshift of 7.5.
We now check if it is possible to have constraints at other reionization redshifts, and how our results would change when using light-cone instead of co-eval boxes. 
We perform these tests on one (GT) $\cmfast$ simulation with $\adata=1$, as doing the whole pipeline (from the training of the U-net, its predictions to the AP test) is expensive in computation time and disk storage, and the GT case has tighter error-bars, thus making this a more stringent test.

\begin{figure}[h]
\centering
\includegraphics[width=0.5\textwidth]{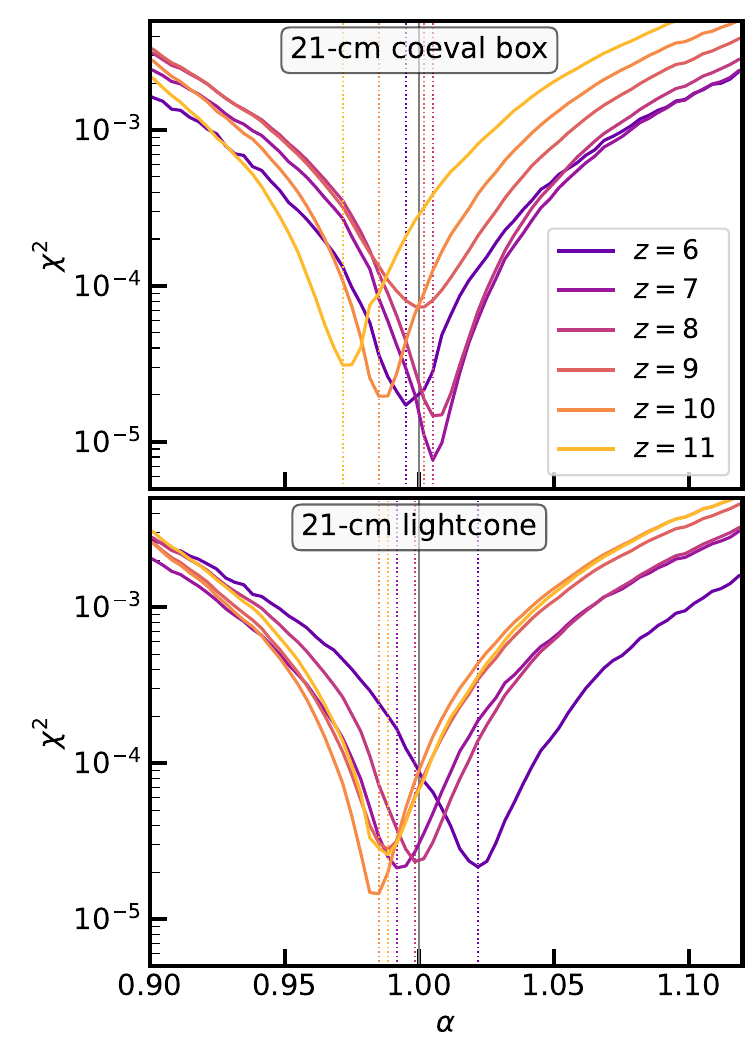}
\caption{$\chi^2$ of GT $\dTb$ bubble stacks. For the top panel, we use a 3D co-eval box, and, for the bottom panel, we use a light cone. The colors represents different redshifts at which we make the AP test. In the light-cone case, the redshift range of the line of sight direction is $\left[ z-0.2,z+0.2\right]$. Only the shape of the $\chi^2$ matters here, and not its amplitude, as we have taken the Covariance matrix to be the identity to avoid running hundreds of simulations.
\label{fig:app_z_lc}}
\end{figure}

For the specific reionization model presented in this work, reionization ends near a redshift of 5.5. We thus tested our work with 3D 21-cm signal boxes at $z\in\left[6,11\right]$. We extract the ionized bubbles from the 21-cm signal using the methods described in Sec. \ref{sec:unet_bubbles}, and we construct the bubble stacks with bubbles that have $\rbubbles>5$ $\cMpc$. 
We choose this cut as we work with only one simulation box (and thus a much smaller mock observed volume), so there are not enough $\rbubbles>25$  $\cMpc$ bubbles in the box to obtain a proper stack.
This test is only a proof of concept, and with observed bubble stacks we would use bubbles with $\rbubbles>25$ $\cMpc$ (from a larger number of simulations, also doing more stacks) as in the main text. We have tested that without noise we can take $\rbubbles>5$ $\cMpc$ and still avoid the noisy, less significant bubbles detected by the watershed code.
The top panel of Fig. \ref{fig:app_z_lc} shows the resulting $\chi^2$ of our AP test. We find that as long as we are able to extract enough bubbles (at least a few hundred) from the data, we can put constraints at these redshifts. Even at late times, when bubbles have percolated, our AP test seems to perform well. The precision is, however, slightly poorer (larger $\chi^2$ width), which is probably due to a smaller number of bubbles and possibly to the greater elongation of each individual bubble. For some redshifts, such as $z=10$ or 11, the $\chi^2$ peaks at lower $\alpha$. This may be because at those redshifts, there are not enough bubbles, so the AP test is less precise, or it could simply be because we use only one simulation and we lack statistics.

As a final test, we have simulated light-cones with $\cmfast$, where the third dimension of the resulting 21-cm boxes represents the redshifts, while the other two directions represent the signal on the plane of the sky. With the reionization model considered here, we chop the third dimension of the light cones to the following ranges: $\left[ z_{\text{mid}}-0.2,z_{\text{mid}}+0.2\right]$ with $z_{\text{mid}}\in\{6,7,8,9,10,11\}$. We then extract the ionized bubbles the same way we do for 3D co-eval boxes (also relaxing the $\rbubbles>25$ $\cMpc$ condition for the same reason as above), and construct the bubble stacks. 
After running them through our AP test, and looking at the $\chi^2$ of the bottom panel of Fig. \ref{fig:app_z_lc}, we can conclude that putting constraints on $\Da H$ is as feasible with light-cone as with co-eval boxes; any systematic shift from the light-cone effect is below the few percent level. This test on light-cones also particularly shows that the RSDs (modeled by default with $\cmfast$) do not alter our result within the $\sim 2\%$ precision we hope to achieve.

\newpage
\section{Varying astrophysical parameters for the bubbles recovery with the U-net} \label{app:varying_params}

\begin{figure}[H]
\centering
\includegraphics[width=0.45\textwidth]{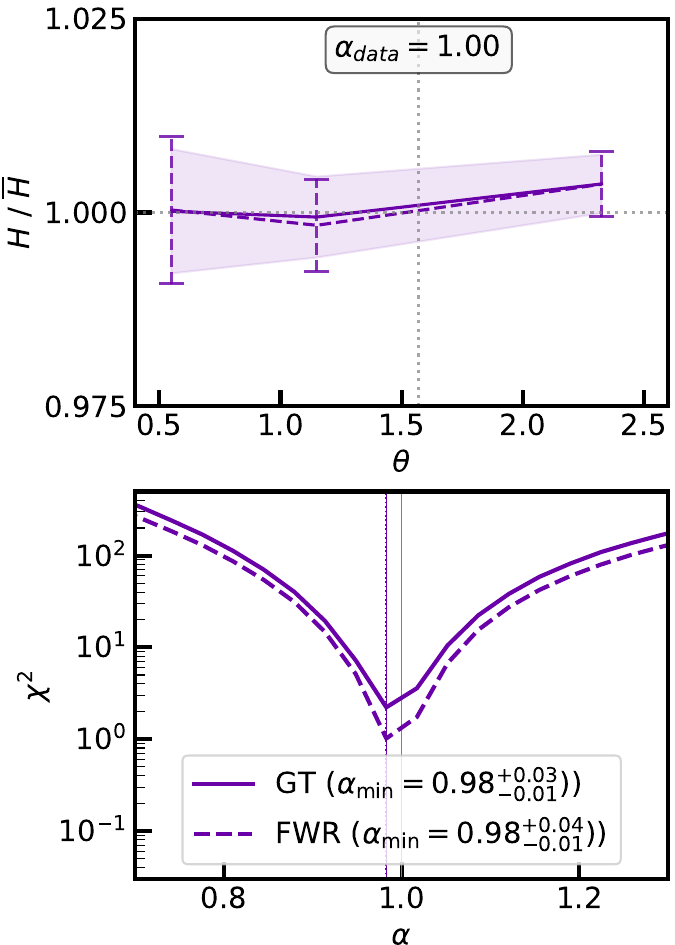}
\caption{AP test results in the case when the U-net is trained on varied parameters (with the prediction still done on the same reionization model as in the text). The top panel shows the ratio of histograms for the best-fit $\alpha$ of the model according to the $\chi^2$ ($\amin$), and the bottom panel shows the $\chi^2$. Both panels display the GT (full lines) and FWR (dashed lines) for an $\adata=1$ bubble stack. The bubble stack is composed of bubbles with $\rbubbles>25$ $\cMpc$ from 25 simulations as in the text, but we only use a 100 bubble stacks to compute the covariance matrix of Eq. \ref{eq:chi2cov}. The error bars are at $1-\sigma$. 
\label{fig:unet_vp}}
\end{figure}

\begin{table*}[]
    \centering
    \normalsize
    \setlength{\tabcolsep}{0.25cm}
    \begin{tabular}{r|l}
    \hline \hline \\[-0.25cm]
    \textbf{Parameters} & \textbf{Description} \\ [0.15cm]
    \hline \\[-0.25cm]
    $f_{*,10}$ & Fraction of galactic gas in stars for $10^{10}$ solar mass haloes \\ [0.15cm]
    $\alpha_{*}$ & Power-law index  of fraction of galactic gas in stars as a function of halo mass \\ [0.15cm]
    \multirow{2}{*}{$\alpha_{*,\text{mini}}$} & Power-law index of fraction of galactic gas in stars as a function of halo mass \\ [0cm] 
    & for molecular cooling galaxies \\ [0.15cm]
    $f_{\text{esc}}$ & Fraction of ionizing photons escaping into the IGM \\ [0.15cm]
    $\alpha_{\text{esc}}$ & Power-law index of escape fraction as a function of halo mass \\ [0.15cm]
    $L_\text{X}$ & Specific X-ray luminosity per unit star formation escaping host galaxies \\ [0.15cm]
    $L_{\text{X,mini}}$ & Specific X-ray luminosity per unit star formation escaping host galaxies for minihalos \\ [0.15cm]
    $\nu_{\text{X,thresh}}$ & X-ray energy threshold for self-absorption by host galaxies (in eV) \\ [0.15cm]
    \hline
    \end{tabular}
    \caption{Astrophysical parameters used to train the U-net on different reionization models.}
    \label{tab:21cmfast_parameters}
\end{table*}

The neural network we use in the main text was trained on only one reionization model, making it potentially model-dependent. In this appendix, we want to circumvent such a dependence. 
To do so, we use the same U-net as in the text \citep{Kennedy2024}, but trained on a different set of simulations. With $\cmfast$, we generate multiple simulations with different sets of astrophysical parameters. The parameters we vary are listed in Table \ref{tab:21cmfast_parameters}, and the values adopted follow the chains of Fig. 8 from \citet{Lazare2024}, except $f_{*,7}$, which was kept fixed for compatibility reasons with the particular version of $\cmfast$ used here. This choice of parameters is, therefore, conservative as \citet{Lazare2024} take into account current upper limits and make no assumption for potential constraints from future surveys.
As the resulting models have different reionization histories, we also use boxes at random redshifts within $\{7,7.25,7.5,7.75,8\}$.
We then train again the neural network with these simulations. The idea is to make the neural network as agnostic as possible to the reionization model used to train it. The resulting predictions from this U-net should therefore be less sensitive to precise parameter choices.

We use this newly trained U-net to predict a new set of GT and FWR $\xHI$ boxes. The reionization model used to make these new predictions is the same as the one used in text for consistency (default astrophysical parameters of $\cmfast$). For this test, we only do the AP test for a hundred GT and noiseless (FWR) bubble stacks with an $\adata=1$ deformation because doing more would be too expensive in terms of computation time and would require a lot of disk storage. With these U-net reconstructions of $\xHI$, we extract the $\rbubbles>25$ $\cMpc$ ionized bubbles and create 100 bubble stacks (each made of bubbles from 25 simulations), before performing the AP test, as in the main text.

Fig. \ref{fig:unet_vp} shows the resulting histogram ratios $H/\bar{H}$ on the top panel and the $\chi^2$ on the bottom panel. The histogram ratios of the top panel are shown for the inferred deformation $\amin$  that minimizes $\chi^2$, in which case they are close to unity.
To compute the $\chi^2$ of the bottom panel, we use Eq. \ref{eq:chi2cov} with a covariance matrix made from 100 bubble stacks. The optimal $\alpha$ deformation given by the $\chi^2$ for both the GT and FWR bubble stack are the following:
\begin{equation}
    \amin^{\text{GT}} = 0.98^{+0.03}_{-0.01} \quad \text{and} \quad
    \amin^{\text{FWR}} = 0.98^{+0.04}_{-0.01}.
\label{eq:app_vp_forecasts}
\end{equation}
We can, therefore, conclude that a model-agnostic U-net can still recover the $x_{\rm HI}$ maps well enough to perform our AP test. We obtain slightly larger error bars here because we use fewer bubble stacks (100) compared to the main text (400). We have not tested the noisy case here for computational reasons, but we expect it to perform as well, as the errors are larger in this case than in the GT and FWR cases.

\newpage
\section{Recalibrating the watershed algorithm in terms of the $\alpha$ deformation} \label{app:watershed_appendix}

\begin{figure}[]
\centering
\includegraphics[width=0.5\textwidth]{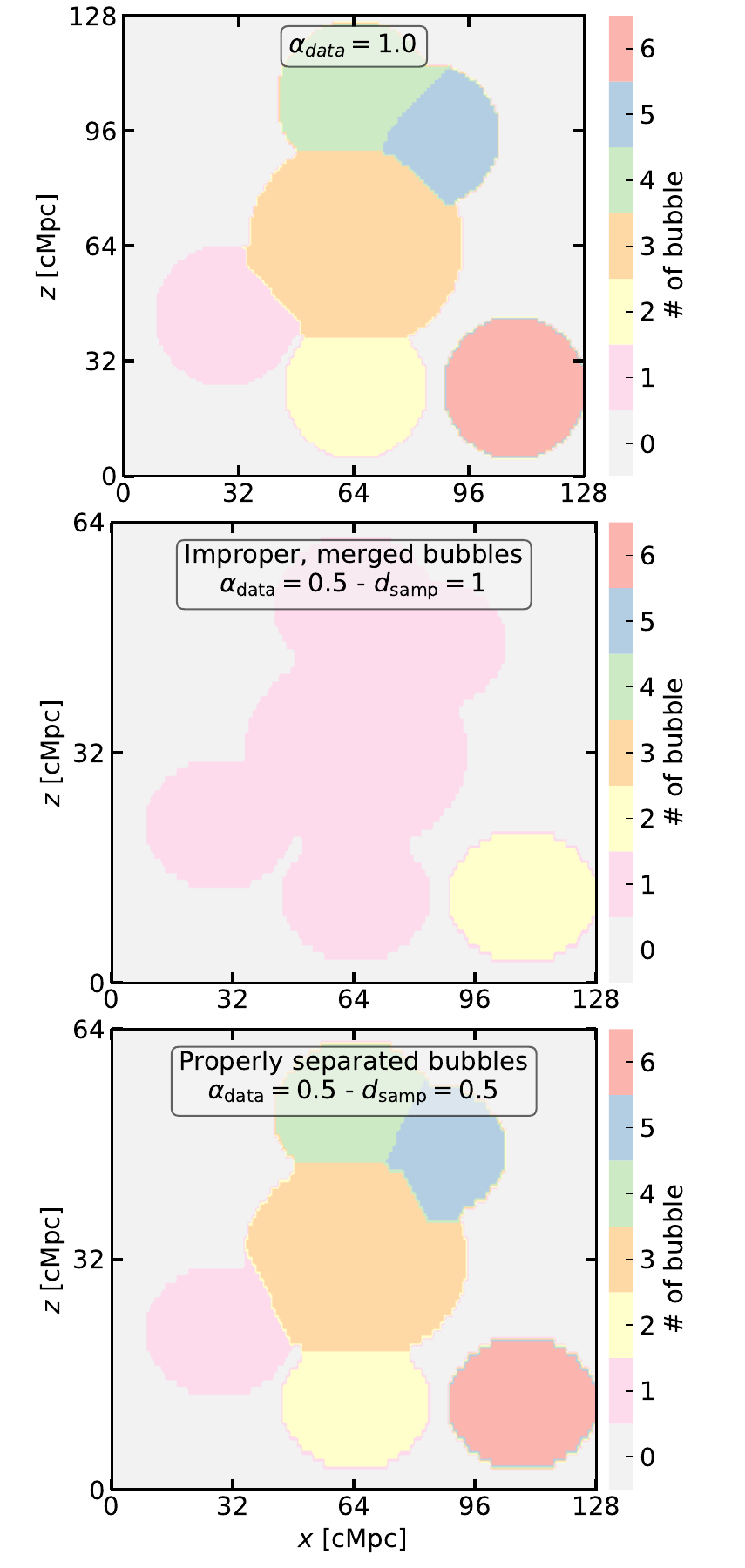}
\caption{Watershed segmentations of toy boxes, which are made up by randomly putting within the box spheres of random sizes that represent the ionized bubbles. The colors are the number associated to the bubbles in the segmentation. The top panel is the regular case with no deformation ($\adata=1$). The middle panel is a slice of the segmentation of a $\adata=0.5$ box, for which the we input $\dsamp=1$ in the watershed code (meaning we do not take into account a deformation of the box). For the bottom panel, we also show a slice of the segmentation of a $\adata=0.5$ box, but, for this one, we input $\dsamp=0.5$ in the watershed code. 
\label{fig:app_watershed}}
\end{figure}

In this Appendix we emphasize that although the watershed algorithm is able to retrieve well ionized bubbles in non-deformed binarized $\xHI$ maps, it fails in deformed boxes. To illustrate this issue, we create a toy box filled with spherical bubbles that are placed randomly, and we use the watershed algorithm to extract the bubbles. The resulting segmentation of this box is shown in the top panel of Fig. \ref{fig:app_watershed}, where the colors simply tag each bubble. Here, we can see that the watershed algorithm identifies each individual bubble as it should. Then, we contract this same toy box with a deformation $\adata=0.5$, and show the resulting segmentation in the middle panel of Fig. \ref{fig:app_watershed}. With this deformed box, the watershed code artificially merges bubbles.
This makes the watershed code extract $\adata\neq1$ bubbles to be more spherical than they should be, biasing the resulting $\alpha$ to be closer to unity in our AP test.

\vspace{5pt}

Ideally, the bubbles extracted from a deformed box should be the same as those extracted from the non-deformed box. That is, they should be located at the same positions and have the same edges (accounting for contraction or elongation). 
In other words, from the watershed code point of view, the $\adata\neq1$ bubbles should be the same as the $\adata=1$ bubbles, but the underlying Euclidean space should be deformed along the $z$ direction. To properly extract those deformed bubbles, we have therefore improved \citet{Lin2016}'s watershed algorithm so that it takes into account potential deformation of the Euclidean space, especially along the $z$ direction. 
As described in Sec. \ref{sec:unet_BS}, this algorithm does a calculation of Euclidean distances on the grid of the input boxes to find the bubbles edge. This is where we need to specify the potential deformation of the $z$ direction: we add a sampling parameter $\dsamp$ that can linearly change the spacing of the grid elements along this direction. Taking $\dsamp=1$ means that the $z$ direction is not deformed. For a given $\adata$ box, this parameter is then directly equal to $\adata$. 
Coming back to our $\adata=0.5$ deformed toy box, we can now extract bubbles with the watershed algorithm using $\dsamp=0.5$. We show the resulting segmentation in the bottom panel of Fig. \ref{fig:app_watershed}. Now, we can correctly identify individual bubbles that correspond exactly to those we extracted from the non-deformed toy box (see the top panel).

In short, in order to properly extract bubbles from any $\adata$ box, we need to let the watershed code know $\adata$. This raises the issue that we do not know $\adata$ a priori, as that is what we want to infer with our AP test. We solve this problem recursively. We start by generating bubble stacks from any $\adata$ dataset with a sampling parameter $\dsamp=1$ (these are the bubble stacks created in Sec. \ref{sec:unet_BS} for instance), and run the AP test to infer $\amin^{(0)}$.
Then, we do the following loop, where $i$ states the number of times the loop is ran:
\begin{itemize}
    \item Generate bubble stacks from the same $\adata$ dataset with $\dsamp=\amin^{(i-1)}$.
    \item Run the AP test to infer $\amin^{(i)}$.
    \item If $\amin^{(i)}$ is within the errors around $\amin^{(i-1)}$, then the inferred value for $\adata$ has converged and the loop can stop. If it is not the case, then the loop goes on.
\end{itemize}
In the main text this looping procedure is implicit, but the inferred $\amin$ values shown in Table \ref{tab:unet_datasets} in Sec. \ref{sec:results_alpha} are the ones obtained thanks to this loop. 
For $\adata=1$, the loop converges directly and the initial inferred value $\amin^{(0)}$ is kept. For any $\adata\neq1$, we find that we only need to run the loop three times to make it converge to our expected $\sim 2\%$ precision, and we keep the last iteration of the inferred $\amin$.


\end{document}